\documentclass[fleqn,usenatbib]{mnras}

\usepackage{newtxtext,newtxmath}
\usepackage{anyfontsize}

\usepackage[T1]{fontenc}
\usepackage{xcolor}
\usepackage{multirow}
\usepackage{graphicx}
\usepackage{subcaption}

\DeclareRobustCommand{\VAN}[3]{#2}
\let\VANthebibliography\thebibliography
\def\thebibliography{\DeclareRobustCommand{\VAN}[3]{##3}\VANthebibliography}

\usepackage{tabularx}

\usepackage{graphicx}	
\usepackage{amsmath}	

\newcommand{\maps}{M\&PS}

\title[Core, chondrite, and icy body]{Measurements of three exo-planetesimal compositions: a planetary core, a chondritic body, and an icy Kuiper belt analogue}

\author[J. T. Williams et al.]{
Jamie T. Williams$^{1}$,\thanks{E-mail: jamietwilliams.astro@gmail.com}
Boris T. G{\"a}nsicke$^{1}$,
Snehalata Sahu$^{1}$,
David J. Wilson$^{2}$,
Detlev Koester$^{3}$,
\newauthor
Andrew M. Buchan$^{1}$,
Odette Toloza$^{4}$,
Yuqi Li$^{5}$,
and Jay Farihi$^{6}$
\\
{1} Department of Physics, University of Warwick, Coventry, CV4 7AL, UK\\
{2} Laboratory for Atmospheric and Space Physics, University of Colorado, Boulder, CO 80303\\
{3} Institut f{\"u}r Theoretische Physik und Astrophysik, University of Kiel, 24098 Kiel, Germany\\
{4} Departamento de Física, Universidad Técnica Federico Santa María, Avenida España 1680, Valparaíso, Chile\\
{5} Institute of Astronomy, University of Cambridge, Madingley Road, Cambridge, CB3 0HA, UK\\
{6} Department of Physics and Astronomy, University College London, London WC1E 6BT, UK}

\date{Accepted XXX. Received YYY; in original form ZZZ}

\pubyear{2025}

\begin{document}
\label{firstpage}
\pagerange{\pageref{firstpage}--\pageref{lastpage}}
\maketitle

\begin{abstract}
The study of planetesimal debris accreted by white dwarfs offers unique insights into the composition of exoplanets. Using far-ultraviolet and optical spectroscopy, we have analysed the composition of planetesimals accreted by three metal enriched H-dominated white dwarfs with effective temperatures of $T_{\mathrm{eff}}\simeq20\,000\,$K. WD\,0059+257 is accreting an object composed of $71.8\pm7.9$\,per cent Fe and Ni by mass, indicating a large core mass fraction of 69\,per cent, similar to that of Mercury. We model this planetesimal as having a differentiated Earth-like composition with 65\,per cent of its mantle stripped, and we find this mass loss can be caused by vaporisation of the planetesimal's mantle during post-main sequence evolution. The tentative S detection in WD\,0059+257 is a possible clue to the nature of the light element in planetary cores, including that of the Earth. The volatile-rich composition of WD\,1943+163 is consistent with accretion of a carbonaceous chondrite-like object, but with an extreme Si depletion. WD\,1953$-$715 accretes a planetesimal which contains $64\pm21\,$per cent of O in the form of ices, likely H$_2$O. This body therefore requires an initial orbit at formation beyond a radial distance of $\gtrsim100$\,au for ice survival into the white dwarf phase. These three planetary enriched white dwarfs provide evidence of differing core fractions, volatile budgets, and initial orbital separations of the accreted planetesimals, all of which help us understand their formation and evolutionary history.
\end{abstract}

\begin{keywords}
white dwarfs -- planets and satellites: interiors -- planets and satellites: composition
\end{keywords}



\section{Introduction}

White dwarfs are the bare cores of stars, formed following the end of nuclear fusion and subsequent ejection of their outer layers. A key characteristic of white dwarfs is their strong surface gravity, $\log\,g\simeq8$, resulting from their mass of $\simeq0.6\,$M$_{\sun}$ contained within an object approximately the size of the Earth. This leads to elements heavier than He rapidly sinking, resulting in a pristine H or He envelope \citep[e.g.][]{Schatzman1949}. Traces of metals in white dwarf atmospheres have been linked to the accretion of planetary material, with $44\pm6\,$per cent of warm (with effective temperatures of $13\,000<T_{\mathrm{eff}}<30\,000$\,K) H-dominated atmosphere white dwarfs showing evidence of active accretion \citep{Koester2014_DAZ,OuldRouis2024}. 

By analysing spectroscopic observations of white dwarfs enriched by planetary material, the composition of the accreted material can be inferred \citep[e.g.][]{Zuckerman2007,Gaensicke2012,Swan2019,Rogers2024b}. Unlike the studies of planets around main sequence stars where the only insight into planetary interior compositions is their bulk density \citep[e.g.][]{Valencia2006} and considering that interior models can be highly degenerate \citep[e.g.][]{Dorn2015}, white dwarfs can reveal the make-up of planets. Investigations into individual planetary enriched white dwarfs have revealed a large diversity, including bulk Earth-like objects \citep[e.g.][]{Klein2010,Xu2014,Doyle2023,Swan2023}, core-rich fragments \citep[e.g.][]{Melis2011,Wilson2015}, mantle and crust-rich fragments \citep[e.g.][]{Zuckerman2011,Hollands2021}, and volatile-rich Kuiper Belt-like objects \citep[e.g.][]{Xu2017,Johnson2022}. This method has been used to find evidence of rocky planetesimal formation at a variety of radial locations \citep{Harrison2018} and to infer the occurrence rate of differentiation\footnote{Defined as the segregation of chemically distinct rock phases into layers, in the case of exoplanets into a core, mantle, and crust.} among accreted planetesimals \citep{Bonsor2020}. 

Photospheric abundances can only directly reflect the composition of the parent body if the accretion rate varies on timescales longer than the one which metals diffuse out of the atmosphere. With diffusion times on the scale of days, this is always true for warm ($T_\mathrm{eff}\gtrsim15\,000\,$K), H-dominated atmosphere white dwarfs \citep{Koester2009_WDs}. Additionally, warm white dwarfs are our only opportunity to simultaneously detect the key rock-forming metals O, Mg, Si, and Fe \citep{Williams2024} which comprise 95\,per cent of the mass of bulk Earth \citep{Lodders2003}. Here, we present the analysis of three such stars (WD\,0059+257, WD\,1943+163, and WD\,1953$-$715) using high-resolution optical and medium-resolution far-ultraviolet (FUV) spectroscopy. These three white dwarfs were identified as being actively accreting planetary debris because of their large photospheric C and Si abundances \citep{Koester2014_DAZ}. Additional measurements of their O abundance implied that their C/O ratios were consistent with Solar System objects \citep{Wilson2016}. In this paper, we present the full complement of detected metals and use these to interpret the nature of the accreted planetesimals. 

\section{Observations} \label{sec:obs}

\begin{table*}
    \caption{Observations log for spectra used in this paper.}
    \centering
    \begin{tabular}{lrrrrrrr}
    \hline
    Name & Date & Instrument & Start Time & Exposure Time (s) & Wavelength Range (\AA) & S/N & Resolving power ($\lambda/\Delta\lambda$) \\
    \hline
    \multirow{3}{*}{WD\,0059+257} & 2011-01-22 & \textit{HST}/COS & 14:39:13 & 1140 & 1132--1274, 1291--1433 & 23 & 12\,000--16\,000 \\
    & 2013-07-13 & \multirow{2}{*}{VLT/UVES} & 09:00:45 & 2980 & \multirow{2}{*}{3700--5000} & \multirow{2}{*}{41} & \multirow{2}{*}{$\simeq 40\,000$} \\
    & 2013-08-19 &  & 06:18:45 & 2980 &  & &  \\
    \hline
    \multirow{3}{*}{WD\,1943+163} & 2011-04-03 & \textit{HST}/COS & 00:42:19 & 600 & 1132--1274, 1291--1433 & 35 & 12\,000--16\,000 \\
    & 2013-06-20 & \multirow{2}{*}{VLT/UVES} & 06:20:08 & 2980 & \multirow{2}{*}{3700--5000} & \multirow{2}{*}{87} & \multirow{2}{*}{$\simeq 40\,000$} \\
    & 2013-06-21 &  & 07:36:26 & 2980 &  & &  \\
    \hline
    \multirow{3}{*}{WD\,1953$-$715} & 2011-05-17 & \textit{HST}/COS & 19:19:42 & 1100 & 1132--1274, 1291--1433 & 28 & 12\,000--16\,000 \\
    & 2013-08-02 & \multirow{2}{*}{VLT/UVES} & 06:06:10 & 2980 & \multirow{2}{*}{3700--5000} & \multirow{2}{*}{59} & \multirow{2}{*}{$\simeq 40\,000$} \\
    & 2013-08-13 &  & 00:41:51 & 2980 & & & \\
    \hline
    \end{tabular}
    \label{tab:observations}
\end{table*}

WD\,0059+257, WD\,1943+163 and WD\,1954$-$715 were observed in the FUV with the Cosmic Origins Spectrograph onboard the \textit{Hubble Space Telescope} \citep[\textit{HST}/COS,][]{Green2012_COS} using the G130M grating (Table\,\ref{tab:observations}). The central wavelength used was 1291\,\AA\ which covers the wavelength range 1132--1433\,\AA\ with a gap between 1278 and 1288\,\AA\ due to a space between the detector segments. We masked the Lyman-$\alpha$ airglow emission present in the region 1213.5--1217\,\AA.

Optical spectroscopy of the three stars was performed using the Ultraviolet and Visual Echelle Spectrograph mounted on the Very Large Telescope \citep[VLT/UVES,][]{Dekker2000_UVES}. These observations were reduced using the {\sc reflex} reduction tool developed by ESO \citep{Freudling2013_REFLEX}. Telluric corrections were applied using the {\sc molecfit} package \citep{Smette2015_MOLECFIT,Kausch2015_MOLECFIT} and the individual exposures were co-added to a single spectrum. The blue arm UVES spectra (3760--4990\,\AA) contains all metal lines detected in the optical. We inspected the red arm UVES spectra (5700--9300\,\AA) for \ion{O}{i}~7774\,\AA\ absorption and \ion{Ca}{ii} infrared triplet emission from the gaseous component of a debris disc \citep[e.g.][]{Gaensicke2006_disc}, but did detect neither feature in any of our targets. No infrared excesses were detected in these three white dwarfs by \textit{Spitzer} \citep{Rocchetto2012}, although the debris may have optically thin dust or an unfavourable inclination which limits its detectable emission \citep{Bonsor2017}.

\section{Methods} \label{sec:num}

\subsection{Photospheric abundances}

\begin{figure*}
    \centering
    \includegraphics[width=0.9\textwidth]{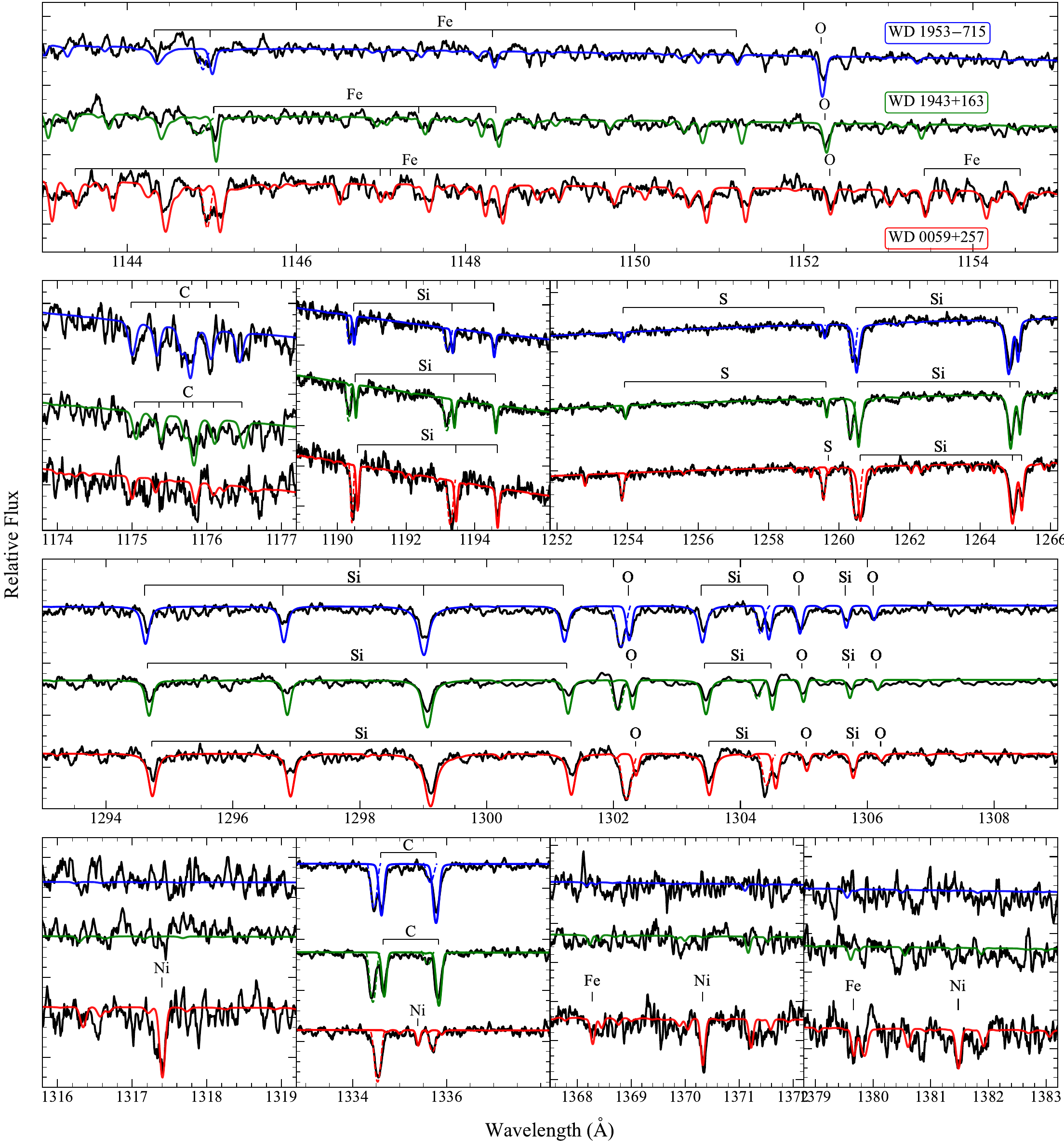}
    \caption{From top to bottom: \textit{HST}/COS spectra (in black, smoothed with a five-point boxcar) of  WD\,1953$-$715, WD\,1943+163 and WD\,0059+257 and the best-fit models in blue, green and red, respectively. The spectra are offset in flux for clarity. The dashes represent which metal lines were used to measure the photospheric abundances. Absorption features in the models that are not marked were not used to fit abundances because they were too weak in the data. The dashed wavelength regions of the models include fits to the ISM lines. The spectra are shown in the observed wavelength frame, hence the photospheric absorption features do not line up with at their laboratory wavelengths. Because the metal abundances are an average of fitting to multiple lines, some lines appear to have a poor fit, particularly the \ion{Si}{III} lines in the $1294-1306\,$\AA\ region. This discrepancy is discussed in Section\,\ref{sec:metals}.}
    \label{fig:COS_spectra}
\end{figure*}

\begin{figure}
    \centering
    \includegraphics[width=0.5\textwidth, keepaspectratio]{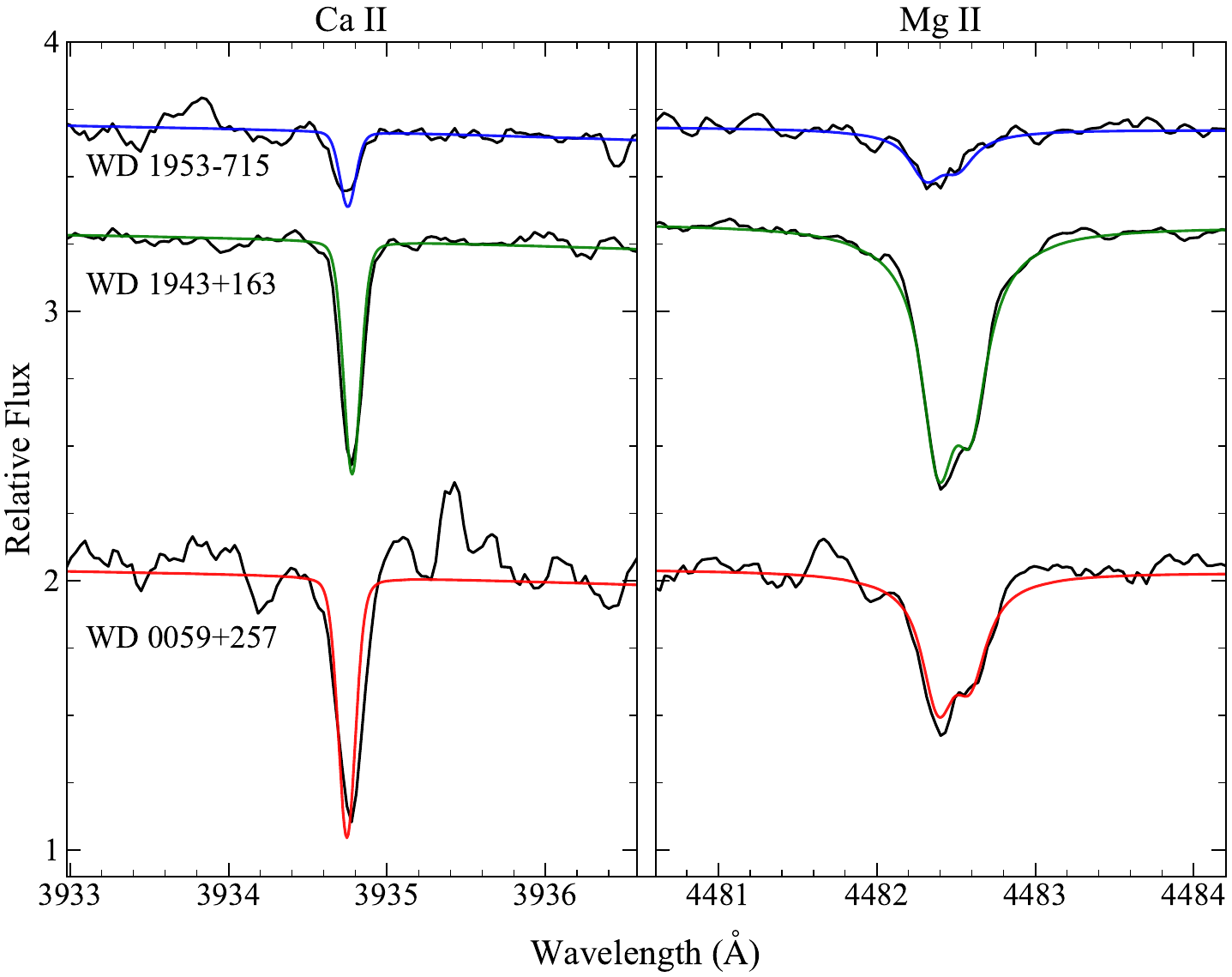}
    \caption{From top to bottom: normalised UVES spectra (in black, smoothed with a five-point boxcar and offset for clarity) of WD\,1953$-$715, WD\,1943+163 and WD\,0059+257 along with the best-fit models in blue, green, and red, respectively showing the \ion{Ca}{ii} (left) and \ion{Mg}{ii} (right) lines. The spectra have been shifted to centre the lines at their vacuum wavelengths.}
    \label{fig:Ca+Mg}
\end{figure}

We detect a wide range of photospheric metals in WD\,0059+257 (O, Mg, Si, S, Ca, Fe, and Ni) as well as in WD\,1943+163 and WD\,1953$-$715 (C, O, Mg, Si, S, Ca, and Fe). To fit the abundances of these elements, and to determine upper limits for N, Al, P, and Cr, we used atmosphere models and synthetic spectra computed with the code described by \citet{koester2010} but with numerous updates and improvements to the equation of state and absorption coefficients. The most important developments are the inclusion of non-ideal effects in the diffusion equation, new calculations for the collision integrals, and significant changes to the diffusion and envelope code \citep{Koester2020}. The absorption lines used in the analysis are summarised in Table\,\ref{tab:lines}. Fixing the $T_\mathrm{eff}$ and $\log\,g$ to the values determined by 
\citet[][see Table\,\ref{tab:DAZ_parameters}]{Sahu2023}, we created a grid of models for each of the elements listed above with abundances in the range $-10<\log\,(\mathrm{Z}/\mathrm{H})<-4$ in steps of 0.25\,dex. The model fluxes were interpolated and scaled to the white dwarf radius and distance (based on inverting the \textit{Gaia}\,DR3 parallax) to find a best match to the observed spectra. When we fit to individual lines, we allowed a small flux offset, normalising around the lines under analysis using a $\chi^2$ minimisation technique. The interstellar lines were also modelled (Section\,\ref{sec:ism}) to obtain accurate metal abundances. Initially, the best-fit abundances were determined for each element using a model grid that contained only lines of that element. Because a number of wavelength regions contain blended lines, we then computed new model grids using the preliminary best-fit metal abundances for all elements and re-fitted the spectra varying the abundance of one element at a time. We continued this iterative process until the abundances converged. The final model fits to the COS spectra are illustrated in Fig.\,\ref{fig:COS_spectra} and to the UVES spectra in Fig.\,\ref{fig:Ca+Mg}, with the resulting photospheric number abundances listed in Table\,\ref{tab:DAZ_abundances}.

\subsubsection{ISM line fitting} \label{sec:ism}
Some of the photospheric metal lines in the FUV are contaminated by absorption in the interstellar medium (ISM) which can affect the abundance measurements. Given the COS velocity resolution of $\simeq15\,\mathrm{km\,s}^{-1}$, these ISM lines may either be resolved or blended with the photospheric lines, depending on the line of sight velocity of the ISM and the systemic velocity and the gravitational redshift of the white dwarf. Hence, to determine accurate abundances, the contribution of the ISM lines must be included in the fits to the photospheric lines. We modelled the ISM lines using Gaussian functions where the free parameters are their amplitudes, standard deviation, and centroids of the velocity, where we restricted the widths to be within the range of $0.01-0.03$\,\AA\ corresponding to a velocity dispersion of $3-7\,\mathrm{km\,s}^{-1}$. This is a reasonable approximation considering the warm temperature of the local ISM ($\simeq7000$\,K, \citealt{seth2004}) and the presence of one to two ISM clouds along the line of sight \citep{malamut2014}. We list the ISM line velocities for each modelled species in Table\,\ref{tab:vism}. To confirm that our fitting of blended ISM and photospheric lines gives reliable abundances, we compare their results with those of uncontaminated lines of the same element. In all cases, we find that their radial velocities and abundances agree.

\subsubsection{Abundance errors}

The errors on abundances were computed using different methods, depending on the number of lines available. For Si, Fe, and Ni, multiple lines were detected in the data, so we fitted each line individually and adopted the weighted mean of individual measurements as the abundance and the standard deviation of the measurements as the error. This method accounts for the statistical error in the fit, systematic uncertainties in the data (via the flux normalisation) and errors in the atomic line data used in the atmosphere models.

Ca and Mg both only have a single line detected in the data, and in the case of C, O, and S the overlap between transitions of different elements required them to be fitted simultaneously. We first obtained the statistical errors in the fits from the covariance matrices but found them to be unrealistically small. To estimate the magnitude of the true error, we use the Planetary Enriched White Dwarf Database\footnote{\href{https://github.com/jamietwilliams/PEWDD}{https://github.com/jamietwilliams/PEWDD}} \citep[PEWDD,][]{Williams2024}. We selected H-dominated atmosphere white dwarfs with $15\,000 < T_{\mathrm{eff}} < 25\,000\,$K and multiple published abundance analyses. The spread in the published abundances are representative of the overall systematic uncertainties in the abundances due to differences in the atmospheric parameters $T_{\mathrm{eff}}$ and $\log\,g$, instruments and models on which these analyses were based, and the fitting procedures that were used. The number of white dwarfs which fulfil these criteria are four for C, O, and S, 13 for Mg, and 16 for Ca, and we use the median of the spread in the abundances of a given element as an estimate of the systematic error, which is added in quadrature to the statistical error.

\subsubsection{Upper limits}\label{sec:upper_lims}

\begin{figure}
    \centering
    \includegraphics[width=0.5\textwidth,height=0.5\textheight,keepaspectratio]{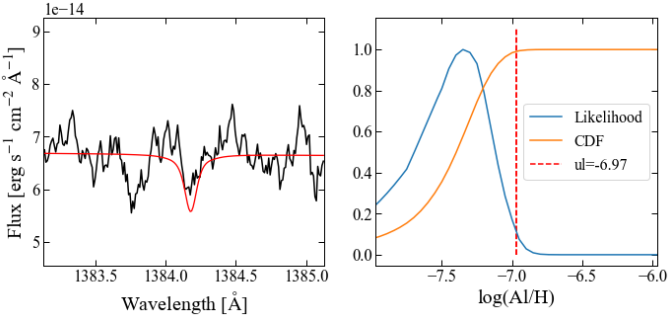}
    \caption{An example illustrating the determination of upper limits on the photospheric abundances. The left panel shows the data around the \ion{Al}{III} line at 1384.1\,\AA\ in black, with a model fit at the Al upper limit overlaid in red. The right panel shows the CDF (orange) of the likelihood (blue) as a function of the Al abundance. We define the upper limit (ul) as being the $99^{\mathrm{th}}$ ($3\sigma$) percentile of the CDF, which is the maximum number abundance possible without returning a statistically significant detection.}
    \label{fig:upper_lim_example}
\end{figure}

We do not detect N, Al, P, or Cr in the atmospheres of any of the three white dwarfs. We also do not detect C in WD\,0059+257 and Ni in WD\,1943+163 and WD\,1953$-$715, however, we have determined upper limits on their abundances following the procedure of \citet{Hollands2020}. We selected the spectral region where a photospheric line is expected and compared it with a grid of synthetic spectra, including the ISM model wherever required. We varied the model abundances by 0.01\,dex and determined the $\chi^{2}$ considering the uncertainties in the observed fluxes. The $\chi^{2}$ values were then converted to likelihood considering Jeffrey's prior and the cumulative distribution function (CDF) was computed. Finally, the abundance at the 99\,per cent confidence interval of the CDF representing the 3$\sigma$ detection threshold was chosen as the upper limit. This method is illustrated in Fig.\,\ref{fig:upper_lim_example} for determining the Al upper limit in WD\,0059+257. 

\subsubsection{Notes on individual elements} \label{sec:metals}



\begin{figure}
    \centering
    \includegraphics[width=\linewidth]{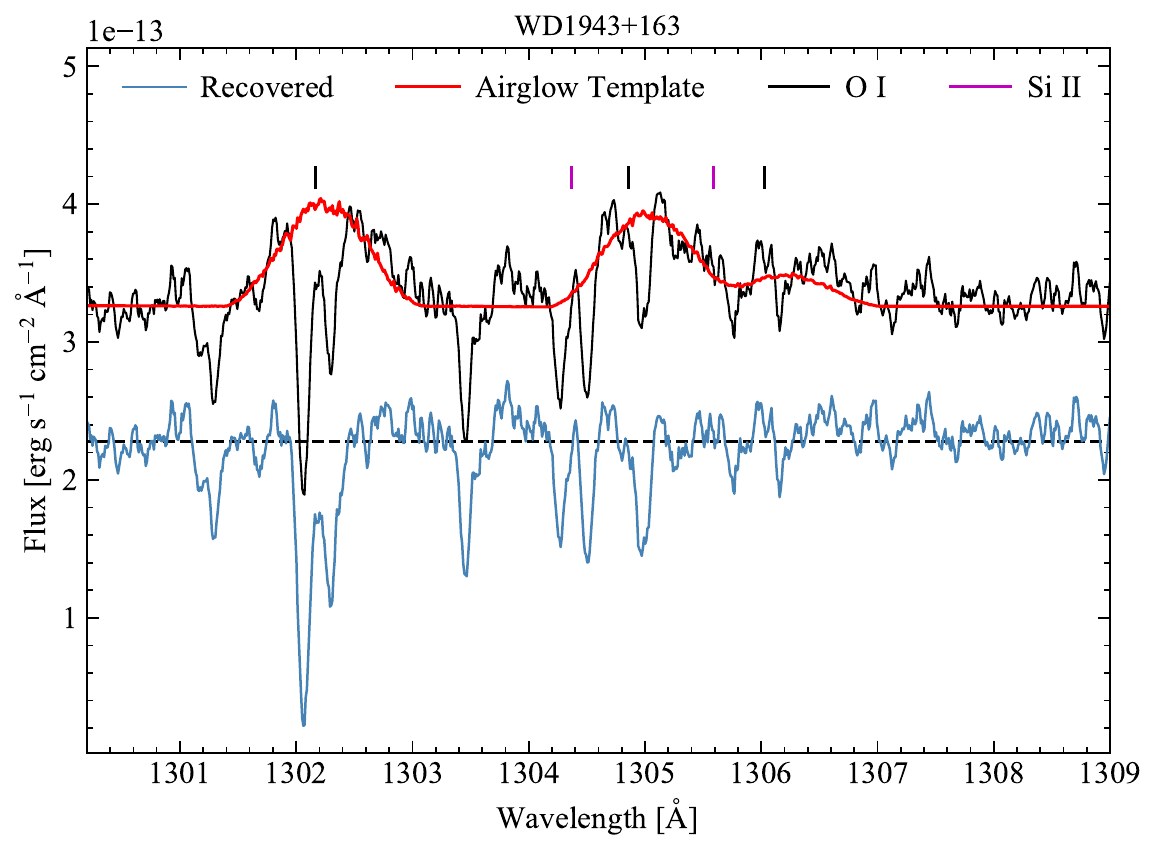}
    \caption{\textit{HST} COS spectrum of  WD\,1943+163 in the region affected by airglow, before (black line) and after (blue line) we apply the airglow correction as described in Section\,\ref{sec:metals}. The best-fit airglow template for the raw spectrum is shown as a solid red line. For clarity, the corrected spectrum is slightly offset downwards in flux, with the dashed black line showing the continuum. Correction for airglow is required for an accurate determination of the abundances of \ion{O}{i} (black dashes) and \ion{Si}{ii} (pink dashes) which have photospheric absorption lines within this wavelength region.}
    \label{fig:airglow_corr}
\end{figure}

The three \ion{O}{i} lines at 1302.2\,\AA, 1304.9\,\AA, and 1306.0\,\AA\ are all affected by O airglow from the Earth's atmosphere to varying degrees, with the former two also affected by ISM lines. Extracting only the data obtained during \textit{HST}'s night-time can remove airglow, but WD\,0059+257 and WD\,1943+163 were only observed during the day-time. Therefore, to reduce the effect of airglow, we used the community-generated templates for \ion{O}{i} lines\footnote{\url{https://www.stsci.edu/HST/instrumentation/cos/calibration/airglow}}  \citep{Bourrier2018}. The templates were obtained with COS G130M grating and a central wavelengths of 1291\,\AA, hence with the same setup as our observations. The velocity and amplitude of the template airglow lines were varied to find a best match to those in our COS data, masking the strong absorption lines from \ion{Si}{ii} and \ion{O}{i} in this process. The original and airglow corrected spectra for WD\,1943+163 are shown in Fig.\,\ref{fig:airglow_corr}. After the airglow correction, we follow the procedure to fit the ISM and photospheric lines together. To confirm that the airglow minimisation and ISM lines did not alter the actual O abundance, we compared the results from regions affected by airglow and ISM to that of the unaffected \ion{O}{i} 1152.2\,\AA\ line and found that the abundances differ by only 0.18, 0.12, and 0.3\,dex for WD\,0059+257, WD\,1943+163, and WD\,1953$-$715, respectively, with the photospheric line velocities being in excellent agreement, within $1\sigma$.

The COS spectra contain many absorption lines of \ion{Si}{II} and \ion{Si}{III} (Table\,\ref{tab:lines}). 
We measured the Si abundance selecting the spectral region 1293--1310\,\AA\ dominated by \ion{Si}{iii} absorption lines and found it to differ by 0.4--0.5\,dex from the regions dominated by \ion{Si}{ii} lines. One of the reasons for the discrepancy could be due to the difficulty in accurately fitting the continuum region (local normalisation) affected by the presence of multiple strong \ion{Si}{iii} lines. The final Si abundance is the weighted average of the abundances determined from the \ion{Si}{ii} and \ion{Si}{iii} lines. In Fig.\,\ref{fig:COS_spectra}, we plot these average abundances, which explains the somewhat lower visual quality of the fits to \ion{Si}{iii} lines. The abundance error we quote adequately accounts for the discrepancy between the \ion{Si}{ii} and \ion{Si}{iii} lines.

We detect strong photospheric \ion{S}{ii} lines in WD\,1943+163 and WD\,1953$-$715 that are well resolved from the ISM lines. 
In the case of WD\,0059+257, we notice a weak photospheric line from \ion{S}{ii} at  $\simeq1259.5$\,\AA. Fitting this line, we found that its  velocity is consistent (within $1\sigma$) with those derived from strong photospheric \ion{Si}{ii} lines, thus confirming the S detection in the atmosphere of WD\,0059+257.

The \textit{HST}/COS detector is affected by fixed pattern noise, which can be mitigated by using all four FP-POS settings. Because the observations analysed here were obtained as part of a snapshot program with strictly limited exposure times, we only used two FP-POS settings. The blue end of the COS spectrum is most strongly affected by the low signal-to-noise ratio (S/N), which explains the apparent visually poor fit of some of the Fe lines in that region.

\subsubsection{Radial velocity}

The radial velocities of the metal lines in the spectra of \textit{HST}/COS and VLT/UVES do not agree within statistical uncertainties, which is particularly apparent for WD\,1953$-$715 (see Table\,\ref{tab:DAZ_abundances}). The dominant source in the systematic uncertainties is the COS wavelength zero-point accuracy\footnote{\href{https://hst-docs.stsci.edu/cosihb/chapter-5-spectroscopy-with-cos}{https://hst-docs.stsci.edu/cosihb/chapter-5-spectroscopy-with-cos}}, 15\,km\,s$^{-1}$, which is sufficient to explain the discrepancy between the velocities measured with the two instruments. In addition, there are slight disagreements between the velocities of different elements detected with the same instrument, for example Fe exhibits lower velocities than the other elements in all three stars. Poor atomic data could result in inaccurate vacuum wavelengths and hence discrepant radial velocities. However, we do not have a satisfactory explanation for this mismatch.

\begin{table*}
    \caption{Stellar properties of the three white dwarfs analysed. Effective temperature ($T_{\mathrm{eff}}$), surface gravity ($\log\,g$), white dwarf mass ($M_{\mathrm{WD}}$), and cooling age ($t_{\mathrm{cool}}$) are taken from \citet{Sahu2023}. The quoted cooling age is the average of values using the Montreal \citep{Bedard2020} and LPCODE \citep{Althaus2013,Camisassa2016,Camisassa2019} sequences, with errors determined using the tables from \citet{Pathak2024}. \textit{Gaia} DR3 ID, parallaxes are taken from \textit{Gaia} DR3 and distances inferred as inverse parallaxes.}
    \centering
    \begin{tabular}{lrrr}
    \hline
    Name & WD\,0059+257 & WD\,1943+163 & WD\,1953$-$715 \\
    \hline
    WD\,J Name & WD\,J010145.76+260242.70 & WD\,J194531.76+162738.77 & WD\,J195838.65-712343.50 \\
    \textit{Gaia} DR3 ID & 307372225052233984 & 1820678800528443776 & 6422236358302245248 \\
    $T_{\mathrm{eff}}$ (K) & 20\,110 $\pm$ 40 & 19\,250 $\pm$ 30 & 19\,250 $\pm$ 60 \\
    log\,($g$ [cm\,s$^{-2}$]) & 7.89 $\pm$ 0.02 & 7.87 $\pm$ 0.01 & 8.12 $\pm$ 0.02 \\
    Parallax (milliarcsec) & 9.20 $\pm$ 0.05 & 21.4 $\pm$ 0.05 & 14.34 $\pm$ 0.03 \\
    Distance (pc) & 108.7 $\pm$ 0.6 & 46.8 $\pm$ 0.1 & 69.5 $\pm$ 0.1 \\
    $M_{\mathrm{WD}}$ (M$_\odot$) & 0.56 $\pm$ 0.01 & 0.55 $\pm$ 0.02 & 0.69 $\pm$ 0.02 \\
    $t_{\mathrm{cool}}$ (Myr) & $52\pm5$ & $62\pm5$ & $106\pm4$ \\
    \hline
    \end{tabular}
    \label{tab:DAZ_parameters}
\end{table*}

\begin{table*}
    \caption{Photospheric number abundances, log\,(Z/H), steady-state accretion rates, $\log\,(\dot{M}_{\mathrm{acc, Z}}$ [g/s]), and photospheric line radial velocities, $v_{\mathrm{Z}}$ of the three white dwarfs studied. The errors on radial velocities are an underestimate, since \textit{HST}/COS has a wavelength accuracy of 15\,km\,s$^{-1}$. The abundances of Mg and Ca were obtained using the ground-based VLT/UVES instrument and hence offset.}
  \begin{tabular}{lrrrrrrrrrr}
  \hline
     &
      \multicolumn{3}{c}{WD\,0059+257} &
      \multicolumn{3}{c}{WD\,1943+163} &
      \multicolumn{3}{c}{WD\,1953$-$715} \\
     Z & log\,(Z/H) & $\log\,\dot{M}_{\mathrm{acc, Z}}$ & $v_{\mathrm{Z}}$ (km\,s$^{-1}$) & log (Z/H) & $\log \dot{M}_{\mathrm{acc, Z}}$ & $v_{\mathrm{Z}}$ (km\,s$^{-1}$) & log\,(Z/H) & $\log\,\dot{M}_{\mathrm{acc, Z}}$ & $v_{\mathrm{Z}}$ (km\,s$^{-1}$) \\
     \hline
     C & $\leq-8.79$ & $\leq3.56$ &  & $-6.37 \pm 0.16$ & 6.05 & $26.1\pm2.3$ &  $-5.75 \pm 0.16$ & 6.80 & $19.7\pm0.1$\\
     N & $\leq-5.98$ & $\leq6.62$ &  & $\leq-6.28$ & $\leq6.44$ &  & $\leq-5.32$ & $\leq7.47$ &  \\
     O & $-5.72\pm0.14$ & 7.03 & $44.2\pm0.6$ & $-5.70\pm0.13$ & 7.28 & $30.2\pm1.7$ & $-5.51\pm 0.19$ & 7.47 & $22.3\pm0.9$ \\
     Mg & $-5.84\pm0.14$ & 6.71 & $46.3\pm1.5$ & $-5.64\pm0.14$ & 6.87 & $31.4\pm1.5$ & $-6.35\pm0.14$ & 6.31 & $9.9\pm1.5$ \\
     Al & $\leq-6.97$ & $\leq5.67$ &  & $\leq-7.15$ & $\leq5.57$ &  & $\leq-6.78$ & $\leq6.04$ &  \\
     Si & $-6.26\pm0.24$ & 6.43 & $47.0\pm3.3$ & $-6.57\pm0.25$ & 6.25 & $32.2\pm0.9$ & $-6.39\pm0.22$ & 6.49 & $25.3\pm2.0$ \\
     P & $\leq-7.70$ & $\leq5.11$ &  & $\leq-8.39$ & $\leq4.64$ &  & $\leq-8.01$ & $\leq5.05$ &  \\
     S & $-7.33\pm0.29$ & 5.60 & $49.8\pm1.0$ & $-6.54\pm 0.10 $ & 6.64 & $31.3\pm0.5$ & $-6.66\pm0.12$ & 6.59 & $27.2\pm0.7$ \\
     Ca & $-6.71 \pm 0.18$ & 6.27 & $42.9\pm1.0$ & $-7.01\pm0.18$ & 5.88 & $35.0\pm0.2$  & $-7.63\pm0.18$ & 5.45 & $6.5\pm1.0$ \\
     Cr & $\leq-6.59$ & $\leq6.61$ &  & $\leq-6.85$ & $\leq6.35$ &  & $\leq-6.54$ & $\leq6.78$ &  \\
     Fe & $-5.54\pm0.16$ & 7.73 & $41.4\pm0.6$ & $-6.22\pm0.17$ & 7.09 & $25.4\pm1.1$ & $-6.43\pm0.19$ & 6.98 & $16.7\pm2.1$ \\
     Ni & $-6.80 \pm 0.24$ & 6.53 & $42.3\pm3.2$ & $\leq-7.60$ & $\leq5.82$ &  & $\leq-7.64$ & $\leq5.85$ &  \\
     $\Sigma$ &  & 7.89 &  &  & 7.67 &  &  & 7.74 &  \\
     \hline
     \label{tab:DAZ_abundances}
  \end{tabular}
\end{table*}

\begin{table}
    \centering
    \caption{ISM line velocities for relevant ions.}
    \begin{tabular}{lrrr}
    \hline
       & 
       \multicolumn{3}{c}{$v_{\mathrm{ISM}}$ (km\,s$^{-1}$)} \\
          Ion & WD\,0059+257 & WD\,1943+163 & WD\,1953$-$715 \\
          \hline
          \ion{C}{II} & $1.8\pm0.5$ 
          & $-28.3\pm0.3$ & $-17.2\pm0.5$\\
          \ion{N}{I} &$7.9\pm2.0$ & $-25.0\pm1.7$ & $-12.1\pm2.6$\\
          \ion{O}{I} & $5.8\pm0.5$
          & $-23.1\pm0.5$ & $-12.4\pm0.6$\\
         \ion{Si}{II} &
         $11.1\pm4.1$ & $-24.5\pm0.8$ & $-12.4\pm1.5$\\
         \ion{S}{II} & $12.4\pm0.6$
          & $-25.3\pm0.5$ & $-9.7\pm0.8$\\
          \ion{Fe}{II} & $1.7\pm1.7$ & $-30.6\pm2.6$  & $-8.2\pm3.9$ \\
          \hline
         \label{tab:vism}
    \end{tabular}
\end{table}

\subsection{Accretion rates}

Because metals sink out of the atmosphere within days and accretion events last $\simeq10^4-10^6\,$yrs \citep{Girven2012,Cunningham2021}, it is a safe assumption that accretion is ongoing and in the steady-state phase. At the $T_\mathrm{eff}$ and $\log g$ of these white dwarfs, radiative levitation cannot sustain the observed metal abundances (see Figs.\,5 and 6 in \citealt{Koester2014_DAZ}) and hence we do not take this effect into account. The diffusion flux for each metal is constant throughout the atmosphere, and we take the observed abundance $X_{\mathrm{acc}}$ to be the approximate value at the Rosseland mean opacity $\tau_{\mathrm{Ross}}=2/3$ from which we calculate the accretion rate $\dot{M}_{\mathrm{acc}}$ using the equation \citep{Koester2014_DAZ}:
\begin{equation}
    X \rho v_{\mathrm{diff}} = X_{\mathrm{acc}} \dot{M}_{\mathrm{acc}}
    \label{eq:acc}
\end{equation}
\noindent where $X$ is the metal mass fraction determined from photospheric abundances, $\rho$ the mass density in the atmosphere, and $v_{\mathrm{diff}}$ the diffusion velocity, both of which are computed from atmospheric models \citep{koester2010}. Accretion rates are listed in Table\,\ref{tab:DAZ_abundances}. All analysis in this paper is under the assumption of steady-state accretion.

\section{Mantle loss} \label{sec:melting}

Enhanced stellar luminosity during post-main sequence evolution can alter the composition of planetesimals. Previous studies indicate that this increased luminosity can induce heating and subsequent melting, which leads to differentiation, albeit only for radii $\lesssim30\,$km and originally orbiting within $\approx2\,$au \citep{Li2024}. Here, we investigate the feasibility of a similar process: direct melting during the post-main sequence phase causing the vaporisation and loss of mantle material, which would result in a core-rich planetesimal. \citet{Vanderburg2015} explored mass loss of a planetesimal orbiting close to a white dwarf, and found the rates were sufficient to explain the dust clouds and accretion onto WD\,1145+017. We expand this treatment over the entire post main sequence evolution (in particular the red giant branch, horizontal branch and asymptotic giant branch phases).

We used MESA stellar evolution models for stars of masses $M_0$ = 1\,M$_{\sun}$, 2\,M$_{\sun}$, and 2.5\,M$_{\sun}$ \citep{Dotter2016,Choi2016,Paxton2011,Paxton2013,Paxton2015,Paxton2018} with a metallicity of $Z=0.02$. During the post main sequence evolution, the star loses mass, causing the orbits of planetesimals\footnote{We refer to planetesimals as objects with masses $\lesssim M_\mathrm{Ceres}$} to widen. We assume that the stellar mass loss is adiabatic and isotropic and that radiation and sublimation perturbations are negligible. We do not consider any additional sources of perturbations such as gravitational interactions with planets in this model. Assuming a circular orbit\footnote{An eccentric orbit leads to a maximum increase in effective temperature by $\approx10\,$per cent for the same semimajor axis. This leads to sublimation occurring $\approx0.3\,$au closer to the star.}, the semimajor axis, $a_{i}$, evolves as:

\begin{equation}
    a_{i} = a_0 \left(\frac{M_0}{M_{i}}\right)
    \label{eq:a_i}
\end{equation}

\noindent where $a_0$ is the initial semimajor axis and $M_{i}$ is the stellar mass at each time step $i$. The equilibrium surface temperature of the planetesimal, $T_{i}$ changes due to orbital evolution and increased stellar luminosity, $L_{i}$:

\begin{equation}
    T_{i} = \left(\frac{L_{i}(1-A)}{16\pi\epsilon\sigma (a_i-R_{i})^2}\right)^{1/4}
    \label{eq:T_i}
\end{equation}

\noindent where $A$ and $\epsilon$ is the albedo and emissivity of the planetesimal, $\sigma$ is the Stefan-Boltzmann constant, and $R_{i}$ is the stellar radius. Removal of material from the planetesimal surface is governed by the Jeans escape parameter, which is the ratio of the escape velocity, $v_{\mathrm{esc}}$ to the thermal velocity, $v_{\mathrm{therm}}$:

\begin{equation}
    \lambda = \left(\frac{v_{\mathrm{esc}}}{v_{\mathrm{therm}}}\right)^2 = \frac{Gm_{i}\mu}{k_{\mathrm{B}}T_{i}r_{i}}
    \label{eq:Jeans}
\end{equation}

\noindent where $m_{i}$ and $r_{i}$ are the instantaneous mass and radius of the planetesimal, $\mu$ the molecular mass of the mineral, $G$ the gravitational constant, and $k_\mathrm{B}$ the Boltzmann constant. Material is removed when $\lambda \lesssim 1$.

We consider two planetesimal structures. The first is a rocky planetesimal, comprised of an Fe core accounting for 25\,per cent of the mass, surrounded by an enstatite (MgSiO$_3$) mantle which is the remainder of the mass\footnote{We do not consider a crust because its mass is negligible compared to the mantle.}. We assume $A=0.1$ and $\epsilon=0.9$ similar to that of Ceres \citep{Li2006}. The mass loss rate per unit area for a given mineral is \citep{Vanderburg2015}:

\begin{equation}
    J = \alpha P_{\mathrm{vap}} \sqrt{\frac{\mu}{2\pi k_{\mathrm{B}}T_{i}}} \,[\mathrm{g}\,\mathrm{s}^{-1}\,\mathrm{cm}^{-2}]
    \label{eq:mass_loss_rock}
\end{equation}

\noindent where $k_{\mathrm{B}}$ is the Boltzmann constant and $\alpha$ is the sticking coefficient. We take $\alpha=0.1$ for enstatite and $\alpha=1$ for iron \citep{vanLieshout2014}. $P_{\mathrm{vap}}$ is the vapour pressure, which is dependent on type of mineral and temperature:

\begin{equation}
    P_{\mathrm{vap}} = 0.1e^{-\mathcal{A}/T + \mathcal{B}}\,[\mathrm{Pa}]
\end{equation}

\noindent with values for enstatite of $\mathcal{A}=68\,908$ and $\mathcal{B}=38.1$ \citep{vanLieshout2014}.

The second planetesimal we consider is a rocky enstatite core which makes up 25\,per cent of the mass surrounded by an icy shell of entirely H$_2$O which is the remaining 75\,per cent of the mass. We take $A=0.5$ and $\epsilon=0.5$ due to increased reflection from the icy surface, which is typical of Kuiper Belt objects \citep{Johnson2015}. For ice, we use a sticking coefficient of $\alpha=1$ \citep{SchallerBrown2007} and values for temperature-dependent vapour pressure $P_{\mathrm{vap}}$ ice using:


\begin{equation}
    P_{\mathrm{vap}} = 10^{\mathcal{A} + \mathcal{B}/T + \mathcal{C}\log T +\mathcal{D}T}\,[\mathrm{Pa}]
\end{equation}

\noindent where the constants for H$_2$O ice are $\mathcal{A}=4.07023$, $\mathcal{B}=-2484.986$, $\mathcal{C}=3.56654$, and $\mathcal{D}=-0.00320981$ \citep{Huebner2006}. We assume that the rotation period is short compared to time-step $\Delta t$ and that the planetesimal is not tidally locked because the tide induced by the star on the planetesimal is so weak that spin-orbit synchronization cannot be reached within the planetesimal's lifetime. The mass loss per time-step is:

\begin{equation}
    \Delta m = J4\pi r^2 \Delta t f_{\mathrm{esc}}
    \label{eq:mass_loss}
\end{equation}

\noindent where $f_{\mathrm{esc}}$ is the fraction of gas particles which can escape the gravitational potential of the planetesimal, using a Maxwell-Boltzmann distribution. This leads to a change in mass of:


\begin{equation}
    m_{\mathrm{i+1}} = m_{i} - \Delta m
    \label{eq:delta_mass}
\end{equation}


\noindent As the mass and radius of the planetesimal decreases, the density will increase because the higher density core will make up a larger fraction of the body. To simplify the radius calculation, we adopt a constant mantle density of $\rho$, with the density of the core a factor $\beta$ larger than the mantle, $\beta\rho$, where $\beta>1$. The core mass, $m_\mathrm{core}=0.25m_{0}$, and radius, $r_\mathrm{core}$, will remain constant at:

\begin{equation}
    r_\mathrm{core}^3 = \frac{3\times0.25m_0}{4\pi \beta\rho} = \frac{3m_0}{16\pi\beta\rho}  
\end{equation}

\noindent However, the radius of the mantle will decrease as mass is lost, $m_0-m_i$. The change in volume can be used to calculate the new planetesimal radius:

\begin{equation}
    \frac{m_{\mathrm{mantle},i}}{\rho} = \frac{0.75m_\mathrm{0} - (m_0-m_i)}{\rho} = \frac{4}{3}\pi(r_{i}^3-r_\mathrm{core}^3)
\end{equation}

\noindent After mass loss, the radius will be:

\begin{equation}
    r_{i}^3 = r_\mathrm{core}^3 + \frac{3(m_i-0.25m_0)}{4\pi\rho} = \frac{3m_0}{16\pi\beta\rho} + \frac{3(m_i-0.25m_0)}{4\pi\rho}
\end{equation}

\noindent which gives:

\begin{equation}
    r_{i} = \left ( \frac{(3-3\beta)m_0+12\beta m_i}{16\pi\beta\rho} \right )^{1/3}
    \label{eq:radius}
\end{equation}

\noindent For the rocky case, the Fe-Ni alloy in the core is approximately twice the density of enstatite in the mantle so we take $\beta=2$. For the icy case, water ice is three times less dense than enstatite so we take $\beta=3$. To calculate the density of the mantle, $\rho$, we use Eq.\,\ref{eq:radius} with $m_0=\mathrm{M}_\mathrm{Ceres}$, $r_i=\mathrm{R}_\mathrm{Ceres}$, and $m_i=m_0$. For $\beta=2$, we calculate $\rho=1850\,$kg\,m$^{-3}$ and for $\beta=3$ we find $\rho=1760\,$kg\,m$^{-3}$. This is not the measured density of enstatite or ice, because this planetesimal would not be solid rock. This density is comparable to the measured density of Ceres \citep{Park2020}, but does not match exactly because Ceres is a much more complex body than this simple two layer model.

We record the total mass loss at the end of the simulation as a fraction of the initial mass. The simulations are run with a grid of planetesimals with initial masses $m_{\mathrm{i=0}}=0.01, 0.1, 1\,$M$_{\mathrm{Ceres}}$, and initial orbits spanning $a_0=1.80-3.54\,$au in 0.01\,au steps for the rocky case and $a_0=68-250\,$au in 2\,au steps for the icy case. As the Jeans escape parameter $\lambda$ scales with planetesimal mass, removal of material is precluded for planetesimals with masses more than a few M$_{\mathrm{Ceres}}$. The internal energy of small planetesimals is incapable of inducing rock-metal differentiation, as shown by bodies in the Solar System \citep{Carry2021}. We therefore do not consider objects with masses $<0.01$\,M$_{\mathrm{Ceres}}$ (corresponding to a radius of $\lesssim100\,$km).


\begin{figure}
    \centering
    \includegraphics[width=0.5\textwidth]{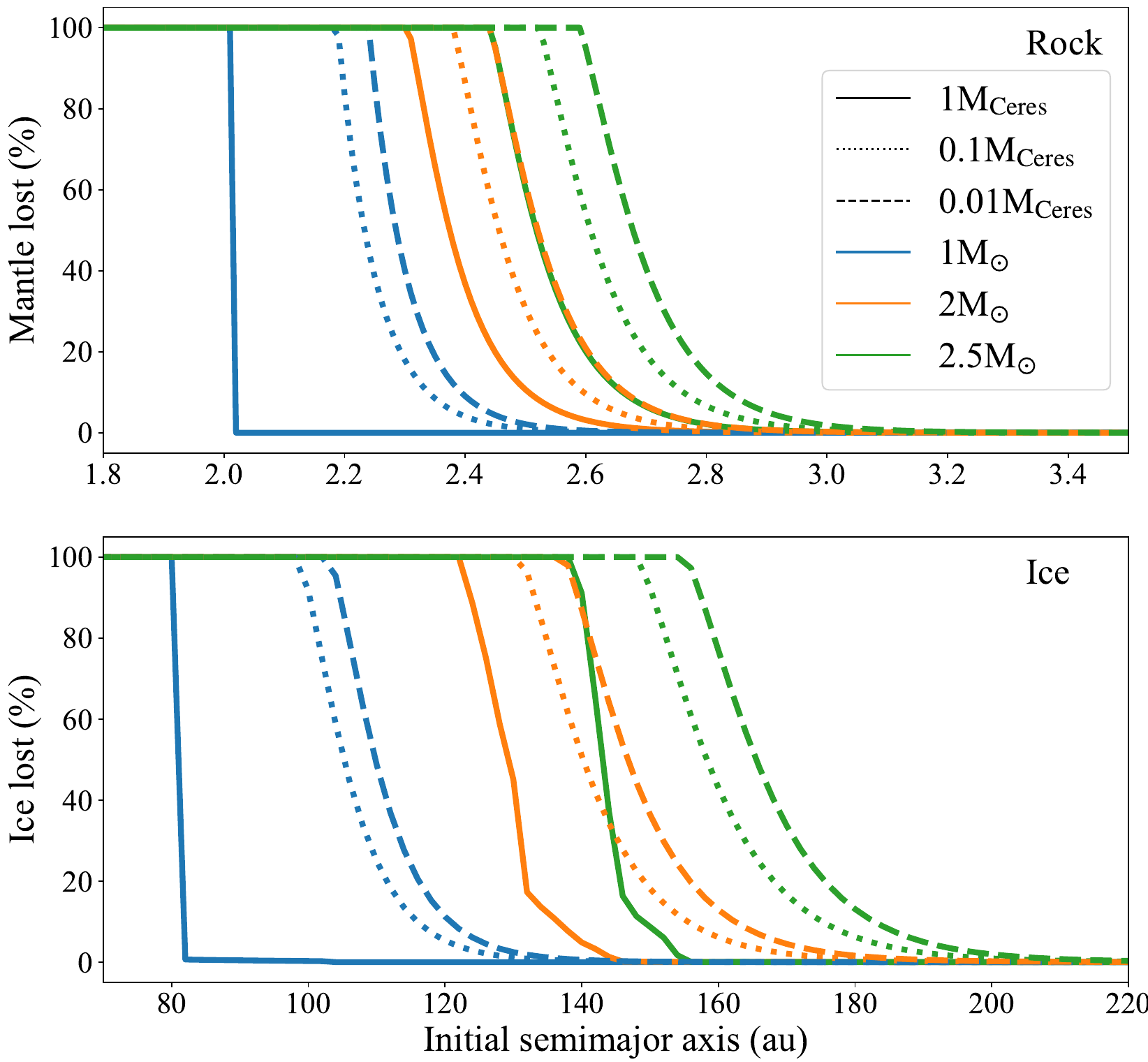}
    \caption{The percentage of mantle lost by a planetesimal with masses $m_{i=0}=1$\,M$_{\mathrm{Ceres}}$, 0.1\,M$_{\mathrm{Ceres}}$, 0.01\,M$_{\mathrm{Ceres}}$ (solid, dotted, and dashed lines respectively) during post-main sequence evolution as a function of initial semimajor axis for stars of initial masses $M_0$ = 1\,M$_{\sun}$, 2\,M$_{\sun}$, and 2.5\,M$_{\sun}$ (blue, orange, green respectively). The top panel considers a rocky MgSiO$_3$ mantle whereas the bottom panel considers an icy H$_2$O mantle.}
    \label{fig:melting}
\end{figure}

The results of our simulations are shown in Fig.\,\ref{fig:melting}, with the sublimation of a rocky and icy mantle in the top and bottom panels respectively. For the rocky (icy) mantle, the entire layer is sublimated within $\simeq2\,$au ($\simeq80\,$au), with smaller planetesimals less resistant to sublimation. Higher mass stars can strip planetary mantles at greater semimajor axes. For all combinations of host star and planetesimal mass, there is negligible rocky (icy) mantle loss exterior to $\simeq3\,$au ($\simeq200\,$au). There exists a range of semimajor axes where partial mantle loss occurs. If a planetesimal belt overlaps with this region, it would serve as a reservoir of objects undergoing this process. This mantle loss means that the sample of objects accreting onto white dwarfs are more core-rich and volatile-poor than their ancestors on the main sequence phase. \citet{Malamud2016} considers a planetesimal with aqueous rock and interior ice and finds that this form of ice can survive closer to the star. In our simulation, we only considered surface ice and found that it is completely sublimated at close distances.

\section{Discussion}  \label{sec:res}

We assume that the photospheric debris in these three white dwarfs originates from a single planetesimal. Simulations of scattering events demonstrate that debris discs will be disrupted by incoming planetesimals, meaning that at any one time, the debris disc will consist of a single object \citep{Jura2008,Mustill2018}. The atmosphere will have no record of previous material accreted more than $\simeq10$ sinking timescales (approximately several months).

\begin{figure}
    \centering
    \includegraphics[width=0.5\textwidth]{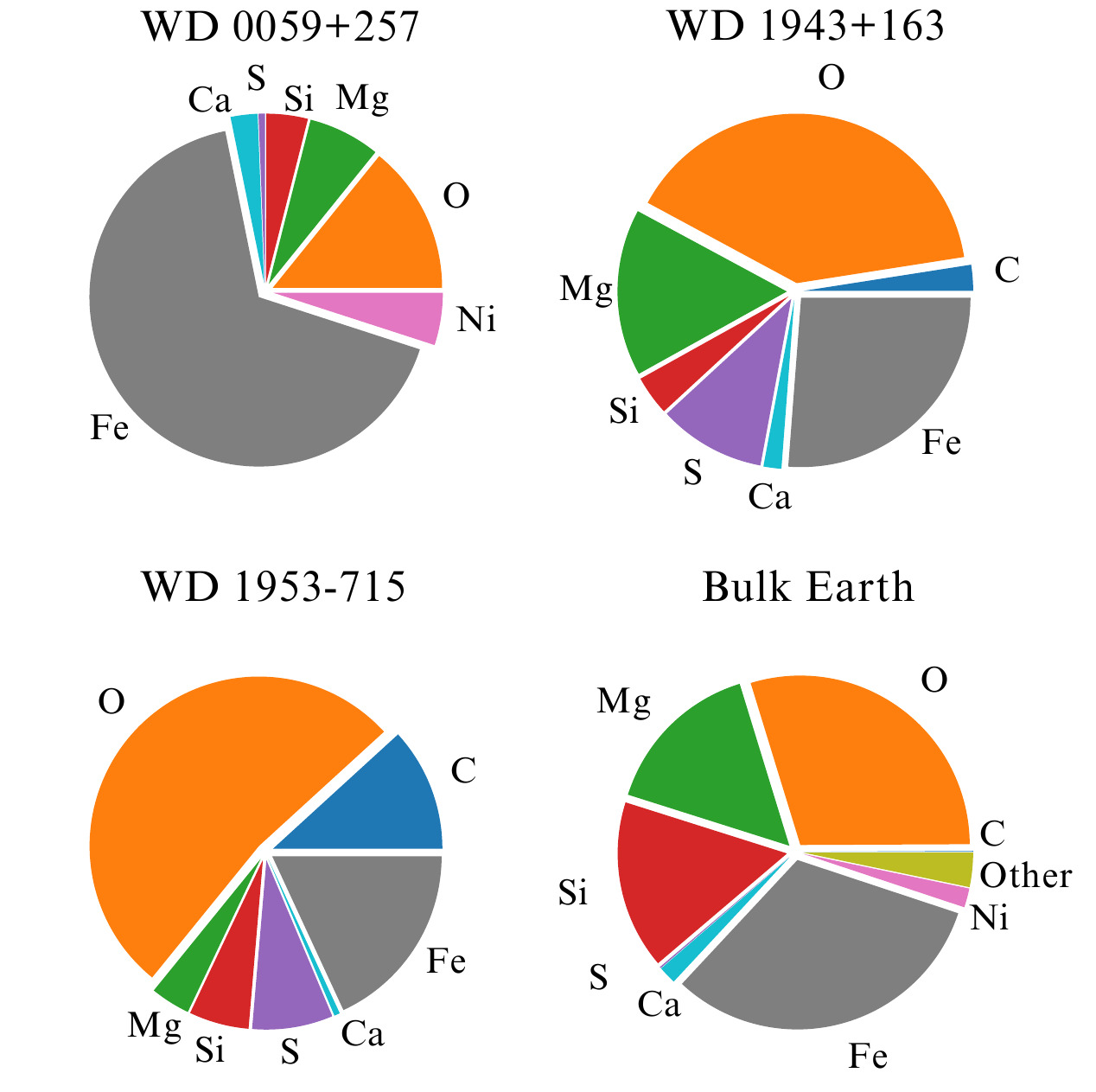}
    \caption{Mass fractions of C, O, Mg, Si, S, Ca, Fe and Ni in the debris accreted by the three white dwarfs. The corresponding mass fraction within the bulk Earth \citep{McDonough2000} are shown as a comparison, with ``Other'' representing the mass fraction of the additional elements in the bulk Earth not detected in the white dwarfs.}
    \label{fig:DAZ_pie}
\end{figure}

We assume that because we detect the major rock-forming elements O, Mg, Si, and Fe, that the total measured accretion rate is within a few per cent of the true value which allows us to compute mass fractions for each metal, listed along with their errors in Table\,\ref{tab:mass_frac} and illustrated in Fig.\,\ref{fig:DAZ_pie}, with bulk Earth for comparison \citep{McDonough2000}. The errors are calculated using a Monte Carlo approach, where we construct Gaussian distributions centred on the metal accretion rates and with a width of their respective error, then randomly select a set of values. We calculate the mass fraction, then repeat this $10^5$ times and take the mean and standard deviation as the final result. This approach is used for all further calculations in this paper.

\begin{table}
    \caption{Metal mass fractions of the accreted material for each white dwarf.}
    \centering
    \begin{tabular}{lrrrr}
        \hline
        & \multicolumn{4}{c}{Mass fraction (\%)} \\
        Z & WD\,0059+257 & WD\,1943+163 & WD\,1953$-$715 & Bulk Earth \\
       \hline
       C &  & $2.5\pm1.0$ & $11.9\pm4.9$ & 0.07 \\
       O & $14.1\pm5.2$ & $39.6\pm8.6$ & $52.2\pm11.6$ & 29.7 \\
       Mg & $6.8\pm2.7$ & $15.8\pm5.0$ & $3.8\pm1.6$ & 15.4 \\
       Si & $4.2\pm3.0$ & $3.8\pm1.1$ & $5.7\pm2.1$ & 16.1 \\
       S & $0.6\pm0.4$ & $10.3\pm5.7$ & $7.7\pm4.2$ & 0.1 \\
       Ca & $2.6\pm1.3$ & $1.7\pm0.8$ & $0.65\pm0.3$ & 1.7 \\
       Fe & $66.8\pm9.0$ & $26.4\pm8.2$ & $18.0\pm7.5$ & 31.9 \\
       Ni & $4.9\pm2.9$ &  &  & 1.8 \\
       \hline
    \end{tabular}
    \label{tab:mass_frac}
\end{table}


\subsection{O budget}

\begin{figure}
    \centering
    \includegraphics[width=0.5\textwidth]{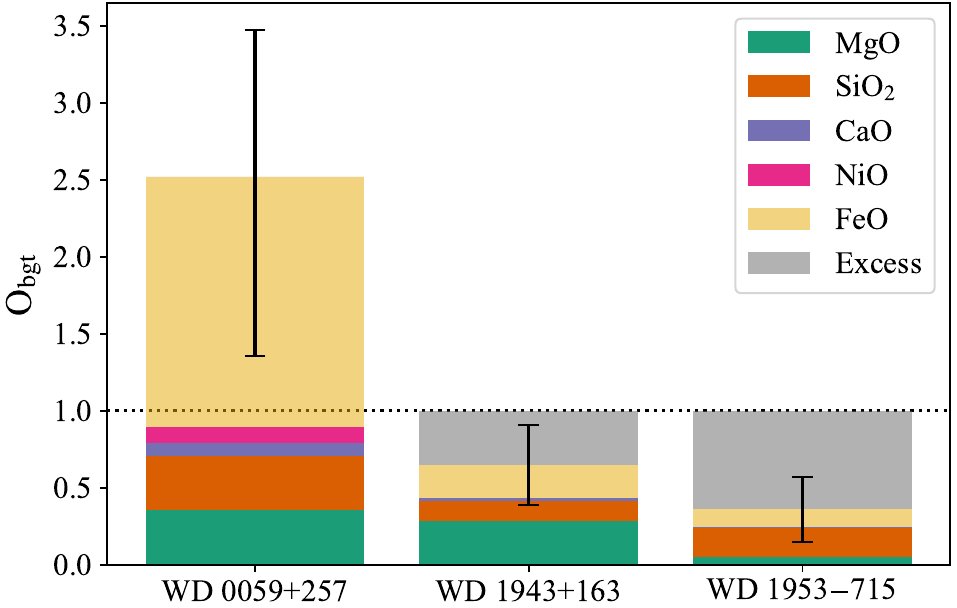}
    \caption{The O$_{\mathrm{bgt}}$ for the debris at the three white dwarfs studied, assuming metals are in the oxides of MgO, SiO$_2$, CaO, NiO, and FeO. We allow the fraction of FeO to vary between fully metallic and fully oxidised. The error bars show the total error on the O$_{\mathrm{bgt}}$. The relative error for all three systems is similar, but WD\,0059+257 has the largest $\mathrm{O}_\mathrm{bgt}$ and hence has the largest absolute error.}
    \label{fig:O_budget}
\end{figure}

All white dwarfs studied in this paper have detections of O, which allows us to calculate their O budget, O$_{\mathrm{bgt}}$, following the prescription of \citet{Klein2010}. We assume a bulk Earth-like oxidation state where the metals are found in the form of the oxides FeO, MgO, SiO$_2$, CaO, and NiO. We can hence write all metal oxides in the form Z$_{a(\mathrm{Z})}$O$_{b(\mathrm{Z})}$, and O$_{\mathrm{bgt}}$ then becomes
\begin{equation}
    \mathrm{O}_{\mathrm{bgt}} = \sum_{\mathrm{Z}} \frac{a(\mathrm{Z})}{b(\mathrm{Z})} \frac{\dot{M}_{\mathrm{Z}}}{\dot{M}_{\mathrm{O}}}\frac{A(\mathrm{O})}{A(\mathrm{Z})}
    \label{eq:O}
\end{equation}
where $\dot{M}$ and $A$ is the accretion rate and atomic mass and the subscript denotes the metal. Strictly speaking O$_{\mathrm{bgt}}>1$ is physically impossible as there is not enough O to account for all the metals in the white dwarf atmosphere. However, we do not know the oxidation state of Fe, which could even be accreted in fully metallic form. FeO and Fe$_2$O$_3$ are possible states in the Earth's crust, with the latter one dominating. However, the state of Fe in the deeper layers is not well established, with volcanic studies suggesting FeO is more common \citep{Basaltic1981,WorkmanHart2005} whereas magnetic studies support a substantial fraction of Fe$_2$O$_3$ \citep{Kupenko2019}. We make a simplifying assumption that FeO is the only oxidised form of Fe and keep the contribution of FeO to the O$_{\mathrm{bgt}}$ as a free parameter. In the case of a balanced O$_{\mathrm{bgt}}=1$ all O can be accounted for by the accreted metal oxides. O$_{\mathrm{bgt}}<1$ suggests an O excess, which requires an additional carrier of O in the accreted debris. The most plausible nature of that extra O carrier is water, $\mathrm{H_2O}$. The O$_{\mathrm{bgt}}$ for each of the white dwarfs is shown in Fig.\,\ref{fig:O_budget}. For WD\,0059+257, most of Fe must be metallic to be consistent with O$_{\mathrm{bgt}}\leq1$, implying the accreted body was in a reduced state. WD\,1943+163 and WD\,1953$-$715 have O excesses, so likely contain ice.

\subsection{Comparison to Solar System objects}

To determine the nature of the material accreted by these three white dwarfs we use a $\chi^2$ minimisation approach to find the most similar Solar System object, using the method outlined in \citet{Xu2013}. We take abundances relative to the combination of the four most common rock-forming elements, O, Mg, Si, and Fe, in order to reduce the sensitivity to the abundance measurement of a single metal:

\begin{equation}
    \chi^2 = \frac{1}{N_\mathrm{Z}}\Sigma \frac{\left[ \frac{\dot{M}_\mathrm{Z}/A_\mathrm{Z}}{\dot{M}_\mathrm{O}/A_\mathrm{O}+\dot{M}_\mathrm{Mg}/A_\mathrm{Mg}+\dot{M}_\mathrm{Si}/A_\mathrm{Si}+\dot{M}_\mathrm{Fe}/A_\mathrm{Fe}} - \left( \frac{\mathrm{Z}}{\mathrm{O}+\mathrm{Mg}+\mathrm{Si}+\mathrm{Fe}} \right)_\mathrm{met} \right]^2}{\dot{M}_\mathrm{Z, err}^2+\dot{M}_\mathrm{O, err}^2+\dot{M}_\mathrm{Mg, err}^2+\dot{M}_\mathrm{Si, err}^2+\dot{M}_\mathrm{Fe, err}^2}
\end{equation}

\noindent where $N_\mathrm{Z}$ is the number of detected metals, the subscript ``met'' denotes the abundance of an individual meteorite, and the subscript ``err'' denotes the error in the accretion rate.

\begin{figure*}
    \centering
    \includegraphics[width=\linewidth]{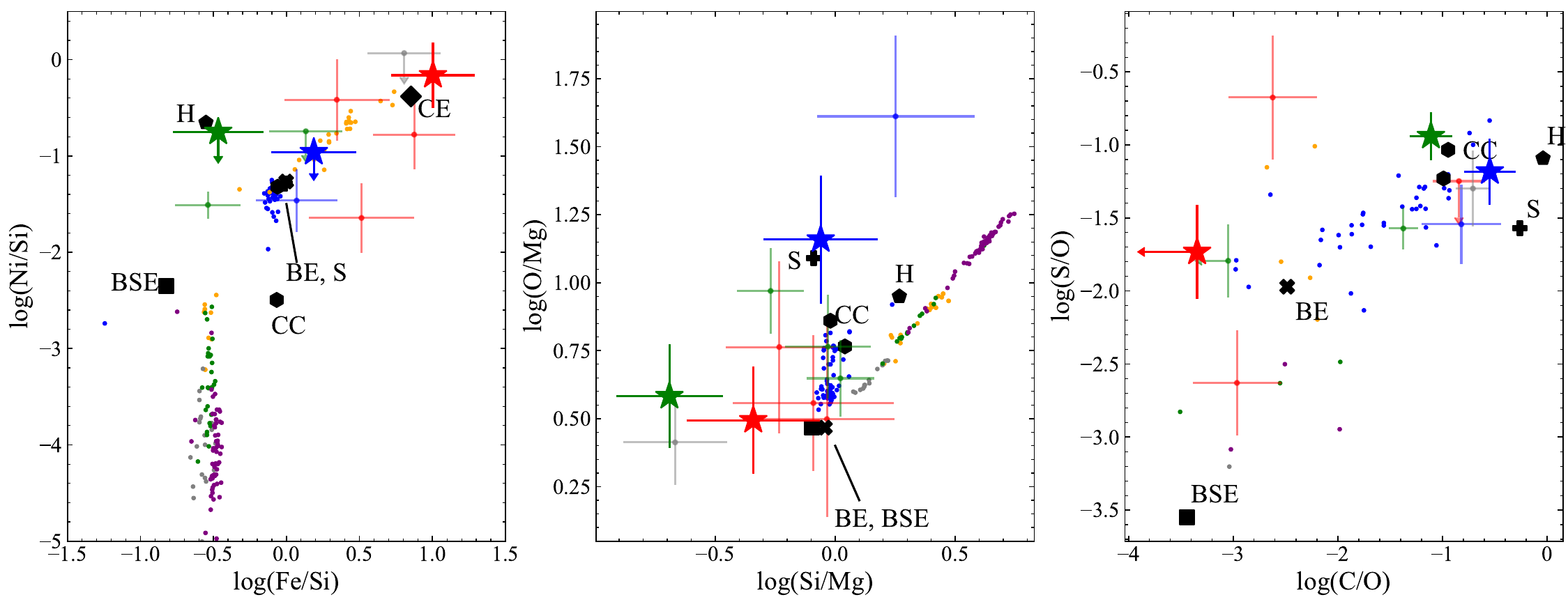}
    \caption{Number abundance ratios generated using PEWDD, with stars representing WD\,0059+257 (red), WD\,1943+163 (green), and WD\,1953$-$715 (blue). For comparison, we include meteorites from \citet{Nittler2004} (carbonaceous chondrites: blue, mesoderites: orange, pallasites: red, diogenites: grey, howardites: green, eucrites: purple). We also show various Solar System objects including bulk Earth (BE, cross; \citealt{McDonough2000}), bulk silicate Earth (BSE, square; \citealt{McDonough2000}), core Earth (CE, diamond; \citealt{McDonough2000}), Halley's comet (H, pentagon; \citealt{Jessberger1988}), the carbonaceous chondrite Tonk (CC, hexagon; \citealt{Nittler2004}), and a solar composition (S, plus; \citealt{Asplund2005}). Using PEWDD, we identify several white dwarfs with comparable temperatures and accreting similar objects to the ones studied in this paper. Core-rich objects are shown in orange (Ton\,345; \citealt{Wilson2015}, PG\,0843+516, and GALEX\,J193156.8+01174; \citealt{Gaensicke2012}), chondritic objects in purple (WD\,J204713.76$-$125908.9; \citealt{Hoskin2020}, GD\,61; \citealt{Farihi2013}, and  WD\,1232+563; \citealt{Xu2019}), WD\,1425+540 which is accreting a Kuiper Belt-like object is shown in pink \citep{Xu2017}, and G\,238-44 which is thought to accrete a mixture of one core-rich and one volatile-rich object is shown in gray \citep{Johnson2022}.}
    \label{fig:num_abun}
\end{figure*}

The material accreted by WD\,0059+257 is most similar to pallasites and mesosiderites \citep{Nittler2004}, both of which are stony-iron meteorites with traces of Fe-Ni alloy. The core Earth composition of \citet{McDonough2000} does not contain data on O or Mg and hence was not fitted in our $\chi^2$ minimisation, but we do note that the composition of WD\,0059+257 closely matches core Earth in the left panel of Fig.\,\ref{fig:num_abun}. For WD\,1943+163, we find the ten best-fitting Solar System objects \citep{Lodders2003,Nittler2004} are all carbonaceous chondrites, with the closest match from the meteorite Tonk, which we illustrate in Fig.\,\ref{fig:num_abun}. The best-fitting Solar system object for WD\,1953$-$715 is with a solar composition, due to the high abundance of volatiles, which are depleted in most rocky objects.

We now discuss each object individually using the O budgets we have calculated and by comparing to Solar System objects. We also explore whether mantle sublimation has affected the composition of each accreted object.

\subsection{\texorpdfstring{WD\,0059+257}{WD 0059+257}}

The planetary debris around WD\,0059+257 is extraordinarily rich in Fe and Ni, which make up a combined $71.8\pm7.9$\,per cent of its mass. Fe and Ni are siderophile elements which preferentially partition into the core during differentiation. Several other white dwarfs share similar siderophile enhancements (GALEX\,J193156.8+01174, \citealt{Melis2011}; PG\,0843+516, \citealt{Gaensicke2012}; Ton 345, \citealt{Wilson2015}; NLTT\,888, \citealt{KawkaVennes2016}; SDSS\,J074153.45+314620.4 and 	SDSS\,J082303.82+054656.1, \citealt{Hollands2018}), with the standard interpretation being that the debris progenitor is a fragment of the core of a chemically differentiated planetesimal. Using \textsc{PyllutedWD}\footnote{\href{https://github.com/andrewmbuchan4/PyllutedWD_Public}{https://github.com/andrewmbuchan4/PyllutedWD\_Public}} \citep{Harrison2018,Harrison2021,Buchan2022}, we find a core mass fraction of 69\,per cent, over double that of Earth's at 32.5\,per cent and comparable to Mercury's of $60-70$\,per cent \citep{Hauck2013}. Mercury is a very reduced object and contains a large mass fraction of Fe, predicted to be due to collisions having stripped the mantle \citep{Benz1988}. The best-fitting meteorite family to the debris, mesosiderites, are speculated to originate from the M-type asteroid Psyche \citep{Davis1999}. This object is hypothesised to be an exposed planetary core \citep{ElkinsTanton2022}, which could have lost its mantle due to a violent collision with another large object \citep{Asphaug2011,Asphaug2017}. 

\begin{figure}
    \centering
    \includegraphics[width=0.5\textwidth]{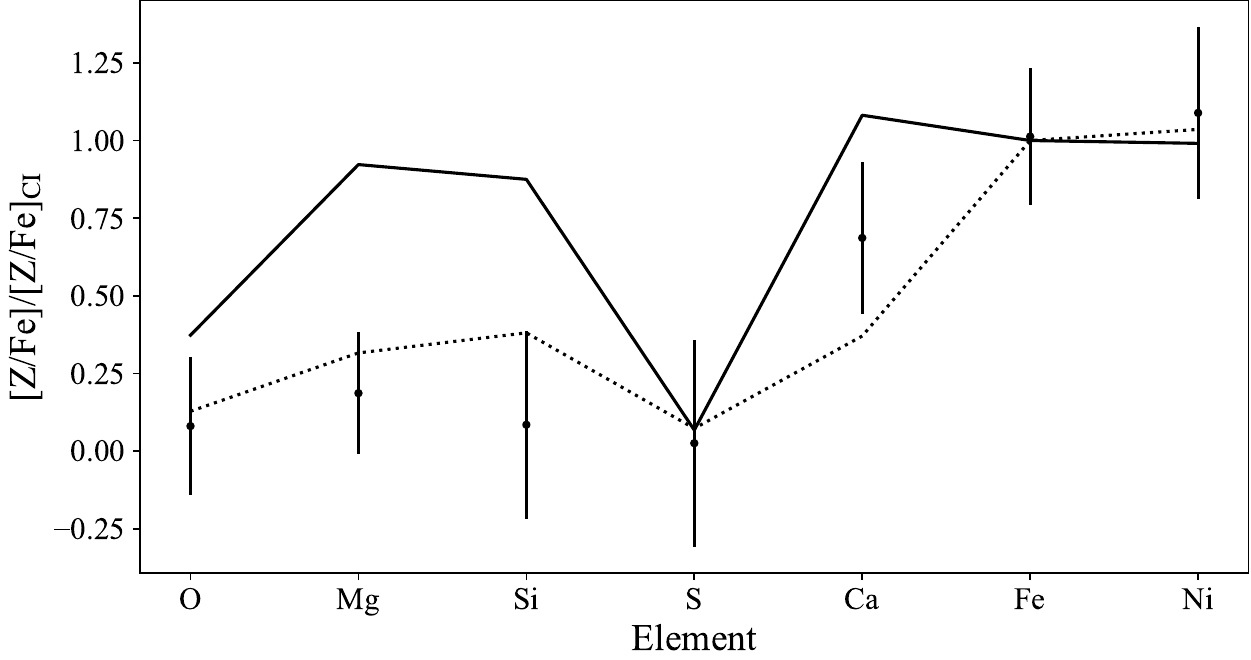}
    \caption{Abundances of the material accreted by WD\,0059+257 relative to Fe and normalised to CI chondrites (black points). These are compared to a bulk Earth model (solid line) and a best-fitting model where 65\,per cent of the mantle is removed (dotted line).}
    \label{fig:sanding}
\end{figure}

However, as discussed in Section\,\ref{sec:melting}, a differentiated object with a mass of $M\lesssim1\,\mathrm{M}_\mathrm{Ceres}$ can lose a portion of its mantle during post-main sequence evolution, making the remaining planetesimal core-rich. With a progenitor mass of $M\lesssim1\,\mathrm{M}_{\sun}$\footnote{The low mass of WD\,0059+257 is in the poorly constrained region of the initial-to-final-mass relation \citep{Heintz2022}, so we can only place an upper limit on the progenitor mass \citep{OuldRouis2024}.}, a planetesimal with an initial semimajor axis in the region of $2.2\lesssim a_0 \lesssim 2.6\,$au could have undergone mantle sublimation. The Fe-rich planetesimals accreting onto GALEX\,J193156.8+011745 \citep{Melis2011} and Ton\,345 \citep{Wilson2015} have compositions that are consistent with a bulk Earth-like differentiated object with a fraction of the mantle removed. Using the prescription from \citet{Melis2011} and compositions from \citet{McDonough2000}, we assume that the core consists of Fe and Ni whereas the mantle contains O, Mg, Si, Ca, Fe, and traces of S and Ni. We gradually remove the mantle so that the composition becomes more core-rich, finding that the best fit is to an object with 65\,per cent of the mantle removed, which is shown in Fig.\,\ref{fig:sanding}.

The possible oxidation state of the parent body suggests that a large fraction of Fe must be in its metallic form, as shown by Fig.\,\ref{fig:O_budget}. The Earth's core consists mainly of a reduced Fe-Ni alloy with $\log\,(\mathrm{Ni/Fe})=-1.235$ \citep{McDonough2000}. The planetesimal accreted by WD\,0059+257 has a similar ratio of $\log\,(\mathrm{Ni/Fe})=-1.2\pm0.3$. Earth's core must also contain a lighter element because the inferred density of the core is 10\,per cent lower than pure Fe-Ni alloy. The light element is commonly predicted to be either O, Si or S \citep{Birch1952,Birch1964,Dreibus1996}. In the object accreted by WD\,0059+257, O and Si have abundances of $\log\,(\mathrm{O/Mg})=0.50\pm0.20$ and $\log\,(\mathrm{Si/Mg})=-0.35\pm0.28$, similar to that of the Earth's mantle of $\log\,(\mathrm{O/Mg})=0.468$ and $\log\,(\mathrm{Si/Mg})=-0.098$. Assuming that this differentiated body underwent mantle stripping similar to the toy model illustrated in Fig.\,\ref{fig:sanding}, this means that O and Si have abundances consistent with being only present in the mantle of this model. However, S is enhanced at $\log\,(\mathrm{S/Mg})=-1.23\pm0.32$ with respect to Earth's mantle at $\log\,(\mathrm{S/Mg})=-3.08$. So, assuming this toy model, S must also be present within the core of the accreted body. We caveat this with the S detection having a large error and the simplicity of our toy model for the planetesimal, but we also note that PG\,0843+516 accretes core-rich material, as well as noticeable amounts of S \citep{Gaensicke2012}.

Whereas the presence of a light element in the core of the Earth is undisputed, determining its nature has proven extraordinarily difficult, as it is impossible to measure the composition of the Earth's core directly. Whereas the number of white dwarfs accreting core-rich objects is still relatively small (left panel of Fig.\,\ref{fig:num_abun}), the identification of several 100\,000 white dwarf candidates using data from the \textit{Gaia} mission \citep{GentileFusilloGaia2021} and the ongoing efforts in spectroscopic follow-up \citep[e.g.][]{Manser2024} demonstrate the potential of unambiguously identifying light elements present within differentiated planetary cores from studying debris-accretion white dwarfs.

\subsection{\texorpdfstring{WD\,1943+163}{WD 1943+163}}


Distinguishing between bulk Earth and chondritic material is possible through observations of the volatile elements C and S in the FUV. Although the refractory elements observed in WD\,1943+163 agree with the composition of the bulk Earth (Figs.\,\ref{fig:DAZ_pie}, \ref{fig:num_abun}), the enhanced volatile abundances from C, O, and S suggest that the accreted material is a carbonaceous chondrite, see the right panel of Fig.\,\ref{fig:num_abun}. 

Carbonaceous chondrites are the primitive building blocks of the Solar System \citep{Alexander2012}, and have not undergone melting or differentiation. They are among the most C-rich objects known \citep{Lodders2003} and are predicted to have formed in the outer Solar System \citep{Wood2005,Warren2011}. The abundances of volatiles within a planetary object are expected to increase with increasing distance from the host star \citep{Larimer1967,Lewis1972} due to lower temperatures during planet formation because of the reduced levels of volatile sublimation. The volatile enhancement in the planetesimal accreted by WD\,1943+163 therefore suggests that it formed at a large separation from its host star and that it did not undergo incomplete accretion \citep{Albarede2009} or volatile outgassing like Earth did \citep{Thompson2021,Thompson2023}. The large radial distance from the star is further supported by the fact that it did not lose volatiles during post-main sequence evolution \citep{Malamud2016}.

In the Solar System, carbonaceous chondrites contain carbonates and minerals modified by water \citep{Bischoff1998,Alexander2015}. Examining the O$_{\mathrm{bgt}}$ for WD\,1943+163 in Fig.\,\ref{fig:O_budget}, we find that even under the assumption that all Fe is in the form of FeO (the mass fraction of metallic Fe in carbonaceous chondrites is low), there is a slight O excess, which can be interpreted as evidence for water. However, the large mass fraction of C means that C-based ice (CO or CO$_2$) could be present instead of water ice.

\begin{figure}
    \centering
    \includegraphics[width=\linewidth]{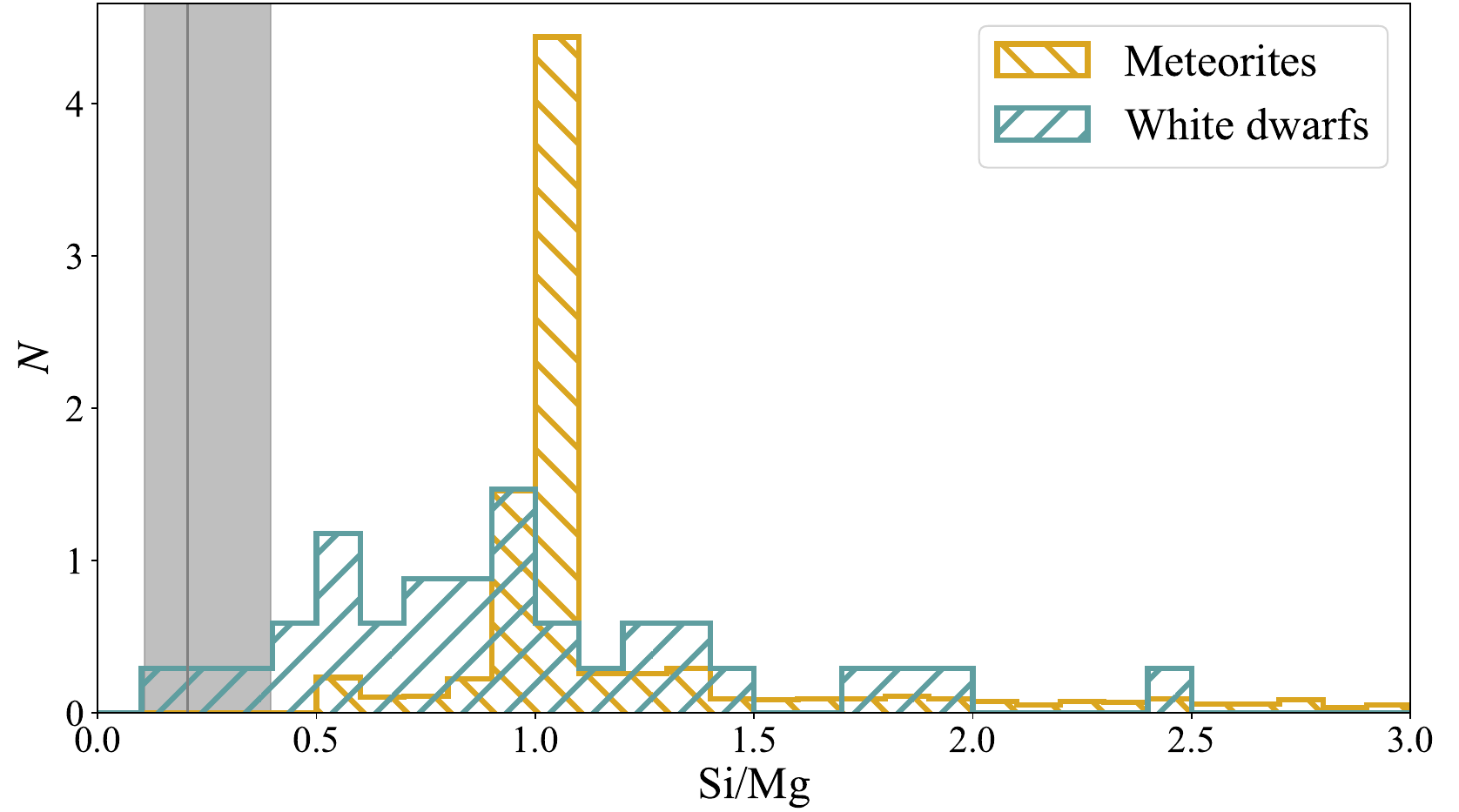}
    \caption{The distribution of Si/Mg abundances for meteorites in the Solar System \citep{Nittler2004} and for white dwarfs in PEWDD with accretion rates for Si and Mg. The vertical line and shaded region is the Si/Mg abundance and error bounds for the material accreted by WD\,1943+163.}
    \label{fig:WD1943_Si}
\end{figure}

Another striking feature of the planetesimal accreted by WD\,1943+163, illustrated in the top right panel of Fig.\,\ref{fig:DAZ_pie} and the centre panel of Fig.\,\ref{fig:num_abun}, is a significant Si depletion. Both chondritic and bulk Earth material have $\mathrm{Si/Mg}\approx1$, whereas the material accreted by WD\,1943+163 has $\mathrm{Si/Mg}\approx0.2$. This cannot be explained by the accretion of a differentiated body such as a mantle-rich fragment as the abundance for bulk silicate Earth is $\mathrm{Si/Mg}\approx-0.8$ \citep{McDonough2000}, mainly consisting of rocks with broadly equal Mg and Si abundances. Furthermore Fig.\,\ref{fig:WD1943_Si} illustrated that no meteorite sample (orange histogram) has an Si/Mg ratio as low as that within the object accreted by WD\,1943+163. The Si discrepancy discussed in Section\,\ref{sec:metals} cannot account for the observed under-abundance, as the adopted Si uncertainties already incorporate this, and a depletion remains even within error.

Using PEWDD, we select 34 white dwarfs that have Mg and Si, and show the resulting Si/Mg ratios as blue histogram in Fig.\,\ref{fig:WD1943_Si}. This distribution, similarly to that of Solar System meteorites, peaks at approximately equal Mg and Si abundances, but has a much longer tail towards Si-depleted objects. Three white dwarfs have Si/Mg abundances that agree with that of the object accreted by WD\,1943+163: PG\,1015+161 \citep{Gaensicke2012}, PG\,1018+411 \citep{Xu2019}, and G\,238-44 \citep{Johnson2022}. All four strongly Si-depleted white dwarfs have warm ($19\,000<T_{\mathrm{eff}}<25\,000\,$K) H-dominated atmospheres and were studied using FUV \textit{HST}/COS spectroscopy. \citet{Johnson2022} concludes that the Si depletion of G\,238-44 cannot be explained by a Si-poor host star. However, Fig.\,\ref{fig:WD1943_Si} shows that strong Si depletion is generally not common among debris-accreting white dwarfs, and the two other white dwarfs studied in this paper, having warm H-dominated atmospheres as well, have $\mathrm{Si/Mg}\approx0.5$ for WD\,0059+257, and Si/Mg$\approx1.5$ for WD\,1953$-$715, consistent with Solar System objects. The geochemical processes that could lead to the Si-depletion detected in the four white dwarfs mentioned above is presently not clear.

\subsection{\texorpdfstring{WD\,1953$-$715}{WD 1953-715}}

Two-thirds of the mass of the object accreted by WD\,1953$-$715 consists of the volatiles C, O, and S (bottom left panel of Fig.\,\ref{fig:DAZ_pie}), and the best-fitting Solar System body is the Sun. Planetesimals that form within the snow line generally consist of rocky, volatile-poor material whereas objects formed further from the star, e.g. those in the Kuiper Belt, are enhanced in volatiles \citep{Morbidelli2000}. Only two Kuiper Belt object (KBO) analogues have been identified accreting onto white dwarfs: WD\,1425+540 \citep{Xu2017} and G\,238-44 \citep{Johnson2022}. The C/O and S/O ratios of WD\,1953$-$715 agree with those measured in these systems, and are close to those of Halley's comet \citep{Jessberger1988}.

We find a lower limit of an O excess of O$_{\mathrm{bgt}}=64\pm21\,$per cent, which is a $3\sigma$ detection. Similar to KBOs, we assume this O is mainly in the form of H$_2$O \citep{Jewitt2004,Filacchione2016}, but note that the large C mass fraction of $11.9\pm4.9\,$per cent suggests C could partially be in the form of CO and CO$_2$ ices, but also molecules such as CH$_4$, HCN, CH$_3$OH, or hydrocarbons \citep{Strom2020}. A Solar System equivalent could be Ceres, which has an icy mantle and is C-rich \citep{Marchi2019}. 

However, the body accreted by WD\,1953$-$715 must have been located further out than Ceres otherwise the icy mantle would have been lost during the evolution of the host star (Fig.\,\ref{fig:melting}). Within the Solar System, the sublimation of H$_2$O of comets begins at 2.5\,au from the Sun \citep{Strom2020}, the result of which are visible as cometary tails. During the post-main sequence evolution, sublimation occurs at much larger separations (bottom panel of Fig.\,\ref{fig:melting}), and the region of complete ice sublimation places constraints on the initial semimajor axes of bodies with thick ice layers accreting onto white dwarfs. Although these simulations consider a simplistic two-layer model with no porosity, we can conclude that the planetesimal accreted by WD\,1953$-$715 likely originated from $\gtrsim150\,$au the host star, which had an initial mass of $M_{\star}\approx2.5$M$_{\sun}$\footnote{We use the initial-to-final mass relation from \citet{Cummings2018}.}. In the context of the Solar System, this would place the body beyond the Kuiper Belt but within the inner radius of the Oort Cloud. Observations of exo-Kuiper Belts suggests that this region is well populated \citep{Chen2006,Lawler2009,Ren2019}.

The strongest evidence of WD\,1425+540 and G\,238-44 accreting KBO analogues is the detection of N, which is strongly depleted in most rocky objects but enhanced in KBOs \citep{Xu2017,Johnson2022}. We did not detect N in WD\,1953$-$715 and could not place a stringent upper limit due to strong ISM line contamination, but this does not rule out a large N mass fraction. However, the necessity of ices to fulfil the O$_{\mathrm{bgt}}$ and the results of the simulation shown in Fig.\,\ref{fig:melting} suggests that this object must have been located at a large radial distance from the host star.

The KBO that accretes onto WD\,1425+540 was likely sent on to a star-grazing orbit by its wide binary companion \citep{Xu2017}. WD\,1953$-$715 does not have a wide binary companion, and \citet{Li2022} finds that trans-Neptunian objects will only have detectable accretion rates of $10^7-10^8\,$g\,s$^{-1}$ for white dwarfs older than 100\,Myr. WD\,1953$-$715 is the oldest of the white dwarfs analysed in this paper with $t_{\mathrm{cool}}=106\pm4\,$Myr, and a total accretion rate of $5.5 \times10^7\,$g\,s$^{-1}$, i.e. within the predicted range. The same delivery mechanism is assumed for the KBO accreting onto G\,238-44 \citep{Johnson2022}, however, there are additional  dynamical pathways for Oort Cloud-analogue objects to be disrupted on to star-grazing orbits \citep{Veras2014c,PhamRein2024}.

\section{Conclusions} \label{sec:con}

Using UV and optical spectroscopy, we measured the bulk abundances of three planetesimals accreting onto white dwarfs, and we discuss their possible structures and evolution. Although the three white dwarfs studied here have similar masses and cooling ages, our analysis shows that the accreted planetesimals display a large diversity their compositions. We demonstrate that sublimation during the post-main sequence phase can alter the properties of planetesimals before they are accreted. 

The planetesimal accreted by WD\,0059+257 shows strong evidence of differentiation, and is among the most core-rich extra-solar planetary body identified so far. In contrast, WD\,1943+163 and WD\,1953$-$715 closely resemble carbonaceous chondrites and Kuiper-belt objects, with the former displaying a substantial level of Si depletion. The detection of volatiles (C, S) are critically important in the interpretation of the formation and structure of planetary bodies, and are only possible though FUV spectroscopy of relatively warm and young white dwarfs. The growing sample of white dwarfs observed with \textit{HST}/COS is beginning to provide detailed statistical insight into the composition of rocky planetesimals in our solar neighbourhood.

\section*{Acknowledgements}

We thank the referee for providing a detailed and constructive report. JTW wants to thank Andrew Swan, Mark Hollands, and Alexander Mustill for useful discussions regarding the interpretation of the accreted debris. This research is based on observations made with the NASA/ESA \textit{Hubble Space Telescope} obtained from the Space Telescope Science Institute, which is operated by the Association of Universities for Research in Astronomy, Inc., under NASA contract NAS 5$-$26555. These observations are associated with the program 12169. Also based on observations made with ESO Telescopes at the La Silla Paranal Observatory under programme ID 091.C$-$0670. This research received funding from the European Research Council under the European Union’s Horizon 2020 research and innovation programme number 101020057 (JTW, BTG, SS). This work has made use of data from the European Space Agency (ESA) mission \textit{Gaia} (\href{https://www.cosmos.esa.int/gaia}{https://www.cosmos.esa.int/gaia}), processed by the Gaia Data Processing and Analysis Consortium (DPAC, \href{https://www.cosmos.esa.int/web/gaia/dpac/consortium}{https://www.cosmos.esa.int/web/gaia/dpac/consortium}). Funding for the DPAC has been provided by national institutions, in particular the institutions participating in the Gaia Multilateral Agreement. This research made use of Astropy\footnote{\href{https://www.astropy.org/}{https://www.astropy.org/}}, a community-developed core Python package for Astronomy \citep{astropy2013,astropy2018}, scipy \citep{scipy2020} and specutils \citep{specutils2023}.

\section*{Data Availability}

The \textit{HST}/COS data used in this paper is available in the MAST archive under the program ID 12169. The VLT/UVES data is available in the ESO archive under program ID 091.C$-$0670.



\bibliographystyle{mnras_}
\bibliography{bibliography} 



\appendix

\section{Line list}

\begin{table}
    \caption{The wavelengths of absorption features that can be found in COS or UVES spectra. Those in italics are contaminated by ISM lines or airglow, which had to be corrected for. The lines in bold were those used in computing the number abundances. For upper limits, the line that gave the tightest constraint is in bold.}
    \centering
    \begin{tabularx}{0.45\textwidth}{cX}
    \hline
        Ion & Vacuum Wavelengths (\AA) \\
        \hline
        \ion{C}{I} & 1158.019,1193.009,1193.240,1261.552,1277.245,1277.282, 1277.550,1277.723,1280.333,1329.578,1355.844,1463.336, 1560.310,1560.683,1561.438,1656.266,1656.928,1657.008, 1657.380,1657.907,1658.122,1930.905 \\
        \ion{C}{II} & 1323.906,1323.951,\textbf{1334.532},\textbf{1335.663},\textbf{1335.708} \\
        \ion{C}{III} & \textbf{1174.930,1175.260,1175.590,1175.710,1175.987,1176.370} \\
        \ion{N}{I} & \textit{1134.165},\textit{1134.415},\textit{1134.980},1167.448,1168.536,1176.510,
        \textit{\textbf{1199.550}},\textit{\textbf{1200.223}},\textit{\textbf{1200.710}},1243.179,1243.306,1310.540, 1319.676,1411.948 \\
        \ion{O}{I} & \textbf{1152.150},\textbf{\textit{1302.170}},\textbf{\textit{1304.860}},\textbf{\textit{1306.030}} \\
        \ion{Mg}{II} & 1369.423,\textbf{4482.383},\textbf{4482.403},\textbf{4482.583},7879.221,7898.539, 8236.899 \\
        \ion{Al}{II} & 1189.185,1190.046,1190.051,1191.808,1191.814,1210.082, 1211.899 \\
        \ion{Al}{III} & \textbf{1379.670},\textbf{1384.132} \\
        \ion{Si}{II} & \textbf{\textit{1190.416}},\textbf{\textit{1193.290}},\textbf{1194.500},\textbf{1197.394},1224.245,1226.798, 1226.979,1227.604,1228.612,1228.739,1229.383,1231.391, 1248.426,1251.164,\textbf{1260.422,1264.738,1265.002,\textit{1304.370}}, \textbf{1305.592},\textbf{1309.276},1350.072 \\
        \ion{Si}{III} & 1206.500,\textbf{1294.545,1296.726,1298.946,1301.149},\textbf{1303.323} \\
        \ion{Si}{IV} & 1393.755,1402.770 \\
        \ion{P}{II} & 1152.818,\textbf{1153.995},1155.014,1156.970,1159.086,1231.184, 1249.830,1310.703 \\
        \ion{P}{III} & 1334.813,\textbf{1344.326} \\
        \ion{S}{I} & 1191.895,1197.548,1425.030 \\
        \ion{S}{II} & 1166.291,1167.512,1233.438,\textit{1250.584},\textbf{\textit{1253.811}},\textbf{\textit{1259.519}} \\
        \ion{S}{III} & 1194.041,1200.956 \\
        \ion{Ca}{II} & \textbf{3934.778},3969.592,8544.435 \\
        \ion{Cr}{II} & 1247.555,1248.219,1398.430,1426.206,1430.853,1431.320, 1431.868, 1432.056,1432.375,1433.007 \\
        \ion{Cr}{III} & 1136.657,1144.089,1146.329,1154.099,\textbf{1161.421}, 1197.341,1201.401,1204.913,1206.370,1209.119,1211.108, 1221.059,1221.887,1225.632,1228.643,1231.857,1232.960, 1236.180,1238.511,1247.831,\textbf{1252.601},\textbf{1259.004},1261.849, 1263.596,1264.197,1269.092 \\
        \ion{Fe}{II} & 1130.443,1133.405,1133.665,1133.675,1138.632,\textbf{1142.312}, 1142.336,1142.474,\textbf{1143.226},1144.049,\textbf{1144.274},\textit{\textbf{1144.939}},
        1146.364,\textbf{1146.833},\textbf{1146.952},\textbf{1147.352},1147.409,1148.072, \textbf{1148.081},\textbf{1148.277},1148.682,1148.945,1148.956,\textbf{1149.615}, 1149.953,1150.289,\textbf{1150.468},\textbf{1150.685},\textbf{1151.146},1152.874, \textbf{1153.272},\textbf{1154.397},1157.631,1159.339,1164.867,\textbf{1166.035}, 1166.974,1168.492,1169.190,\textbf{1170.292},1175.681,1183.829, 1192.030,1198.932,1199.237,1199.672,1200.240,1219.798, 1224.130,1225.496,1228.521,1230.917,1230.927,1260.536, 1260.827,1262.142,1265.641,1266.254,1266.679,\textbf{1267.425}, 1271.986,1272.617,1275.782,1275.822,1277.647,1290.193, 1296.696,1300.047,1309.551,1311.063,\textbf{1358.935},\textbf{1359.055}, 1361.363,\textbf{1361.374},1362.754,1364.579,1366.391,1366.723, \textbf{1368.094},1371.028,1373.718,\textbf{1375.175},1377.986,\textbf{1379.469}, 1383.580,\textbf{1387.218},\textbf{1392.148},1392.817,1397.578,1408.484, 1412.842,1413.706,1417.738,1420.916,1424.717,1430.781, 1430.888,1448.401 \\

        \ion{Fe}{III} & 1130.397,1131.189,1131.908,\textbf{1142.455},1142.950,\textbf{1143.666}, 1214.562 \\
        \ion{Ni}{II} & 1133.730,1134.533,1137.090,1139.638,1140.459,1154.419, \textbf{1164.279},\textbf{1164.575},\textbf{1168.041},\textbf{1171.291},\textbf{1173.477},\textbf{1317.217}, \textbf{1335.201},\textbf{1370.132},\textbf{1381.286},\textbf{1393.324},1411.065 \\
    \hline
    \end{tabularx}
    \label{tab:lines}
\end{table}


\bsp	
\label{lastpage}
\end{document}